\documentclass[journal=./achemso/jpc,manuscript=article,layout=onecolumn]{achemso}
\usepackage{amsmath}
\usepackage[dvipsnames, table]{xcolor}
\usepackage{pdfpages}
\usepackage{amsfonts,amstext,amsgen,amsbsy,amsopn,amssymb}
\usepackage{mathrsfs}

\usepackage{graphicx}
\usepackage{dcolumn}
\usepackage{bm}
\usepackage{physics}
\usepackage{makecell}
\usepackage{multirow}

\usepackage{enumitem}
\usepackage{mathtools}
\usepackage[colorlinks=true, linkcolor=blue, citecolor=blue, urlcolor=blue]{hyperref}
\usepackage{array}

\newcommand{\twqq}[0]{\qquad \qquad} 

\SectionNumbersOn

\title{Tensor-train Thermo-field Memory Kernels for Generalized Quantum Master Equations}

\author{Ningyi Lyu}
\affiliation{Department of Chemistry, Yale University, New Haven, CT 06520, U.S.A.}
\altaffiliation{These authors contributed equally.}
\author{Ellen Mulvihill}
\affiliation{Department of Chemistry, Yale University, New Haven, CT 06520, U.S.A.}
 \altaffiliation{These authors contributed equally.}
\author{Micheline B. Soley}
\affiliation{Department of Chemistry, Yale University, New Haven, CT 06520, U.S.A.} 
\alsoaffiliation{Yale Quantum Institute, Yale University, New Haven, CT 06511, U.S.A.}
\alsoaffiliation{Department of Chemistry, University of Wisconsin-Madison, Madison, WI 53706, U.S.A.}
\author{Eitan Geva}
\affiliation{Department of Chemistry, University of Michigan, Ann Arbor, MI  48109, U.S.A.}
\email{eitan@umich.edu}
\author{Victor S. Batista}
\affiliation{Department of Chemistry, Yale University, New Haven, CT 06520, U.S.A.}
\alsoaffiliation{Yale Quantum Institute, Yale University, New Haven, CT 06511, U.S.A.} 
\email{victor.batista@yale.edu}

\date{\today}

\begin{document}


\begin{abstract}
The generalized quantum master equation (GQME) approach provides a rigorous framework for deriving the exact equation of motion for any subset of electronic reduced density matrix elements (e.g., the diagonal elements). In the context of electronic dynamics, the memory kernel and inhomogeneous term of the GQME introduce the implicit coupling to nuclear motion or dynamics of electronic density matrix elements that are projected out (e.g., the off-diagonal elements), allowing for efficient quantum dynamics simulations. Here, we focus on benchmark quantum simulations of electronic dynamics in a spin-boson model system described by various types of GQMEs. Exact memory kernels and inhomogeneous terms are obtained from short-time quantum-mechanically exact tensor-train thermo-field dynamics (TT-TFD) simulations and are compared with those obtained from an approximate linearized semiclassical method. The TT-TFD memory kernels can provide insights on the main sources of inaccuracies of GQME approaches when combined with approximate input methods and pave the road for development of quantum circuits that could implement GQMEs on digital quantum computers.
\end{abstract}


\section{Introduction}
Quantum dynamics simulations are central to theoretical studies of many areas of chemistry and technological applications, including charge and energy transfer in photosynthetic and photovoltaic systems and a wide range of reactions with nonadiabatic dynamics and photochemical processes, including spin and vibrational energy relaxation as well as polaritonic chemistry.\cite{xu94, ishizaki12, liddell97, liddell02,bredas04,rizzi08, tian11, mishra09, feldt10, zhao12, lee13, lee14a} 
Despite considerable progress over the past few decades, the development of efficient methods for simulations of quantum dynamics remains an outstanding challenge for studies of complex molecular systems at finite temperature.\cite{leggett87, breuer02, nitzan06, weiss12,nakajima58,zwanzig60b,mulvihill22,meyer09,makri99a, jin08, tanimura89, tanimura90, tanimura06, greene17} 
This is primarily due to the computational cost of quantum-mechanically exact simulations, which scales exponentially with the number of degrees of freedom in the system, thereby making such simulations intractable in most complex molecular systems of practical interest. Thus, reduced-dimensionality approaches that can offer more favorable scaling are highly desirable. 

The Nakajima-Zwanzig generalized quantum master equation (GQME)\cite{nakajima58,zwanzig60b} provides a formally exact general-purpose framework for modeling quantum dynamics in reduced dimensionality. It can be obtained for any subset of reduced density matrix elements by using suitable projection operators.\cite{mulvihill22} When focusing on electronic dynamics, the effect of projecting out nuclear degrees of freedoms (DOF) or electronic density matrix elements not included in the subset of interest is accounted for by the {\em memory kernel} and the {\em inhomogeneous term}
 of the GQME. The dimensionality of those spatially and temporally compact quantities is typically much lower than the dimensionality of the overall system since it is determined by the number of reduced density matrix elements included in the subset of interest, allowing for efficient simulations. 

Considerable progress has already been made towards calculating the aforementioned memory kernels and inhomogeneous terms without resorting to perturbation theory.\cite{shi03g, shi04a, zhang06, ka06b, cohen11, wilner13, cohen13a, cohen13b, kelly13, kidon15, pfalzgraff15, montoyacastillo16, kelly15, kelly16, kidon18, pfalzgraff19, mulvihill19a, mulvihill19b, mulvihill21a, mulvihill21b, mulvihill22, xu18, liu18, yan19,  dan22, chatterjee19, brian21}
Much of that progress has been based on the strategy introduced by Shi and Geva,~\cite{shi03g} which relies on formally exact relationships between the memory kernel and the inhomogeneous term and projection-free inputs (PFIs) that are given in terms of two-time correlation functions of the overall system. These PFIs can be obtained from quantum-mechanically exact or approximate (e.g., semiclassical or mixed quantum-classical) input methods.\cite{shi03g, shi04a, zhang06, ka06b, cohen11, wilner13, cohen13a, cohen13b, kelly13, kidon15, pfalzgraff15, montoyacastillo16, kelly15, kelly16, kidon18, pfalzgraff19, mulvihill19a, mulvihill19b, mulvihill21a, mulvihill21b, mulvihill22, xu18, liu18, yan19, brian21, dan22} 

In this paper, we introduce exact memory kernels and inhomogeneous terms obtained from quantum-mechanically exact tensor-train thermo-field dynamics (TT-TFD) simulations.~\cite{Gelin2017,Borrelli2021} To the best of our knowledge, this is the first application of TT-TFD to calculations of memory kernels and inhomogeneous terms of GQMEs. Previously, exact memory kernels have been obtained by the Geva,\cite{shi03g} Shi,\cite{xu18, liu18, yan19, dan22} Makri,\cite{chatterjee19} and Rabani\cite{cohen11, cohen13a, cohen13b, kidon15, ng21} groups. This paper extends the available exact results to include the memory kernels and inhomogeneous terms of the modified GQME and reduced-dimensionality GQME approaches for the spin-boson model. We demonstrate the capabilities of the GQMEs as applied to benchmark simulations of electronic relaxation dynamics in a spin-boson model system, including calculations based on various types of reduced-dimensionality GQMEs. The spin-boson benchmark model provides a useful framework for modeling molecular systems with coupled electronic states. The resulting  quantum-mechanically exact memory kernels and inhomogeneous terms can serve as benchmarks for assessing the accuracy of approximate memory kernels and inhomogeneous terms obtained by approximate input methods. In addition, the reported quantum-mechanically exact memory kernels and inhomogeneous terms could enable the development of quantum circuits for the implementation of GQMEs on digital quantum computers.


The paper is organized as follows. The objectives and scope of our approach are presented in Sec.~\ref{sec:preliminfo}, the GQME formalism 
is outlined in Sec.~\ref{sec:gqme}, and 
the protocol used for calculating the PFIs via TT-TFD is described in Sec.~\ref{sec:tt-tfd}. The utility of combining the GQME and TT-TFD approaches is 
demonstrated for the benchmark spin-boson model
in Sec.~\ref{sec:spin-boson}.
Also included in Sec.~\ref{sec:spin-boson} is a comprehensive comparison between the TT-TFD-based quantum-mechanically exact results and the corresponding approximate results based on PFIs previously obtained with approximate linearized semiclassical mapping Hamiltonian methods.\cite{mulvihill22}
Concluding remarks are provided in Sec.~\ref{sec:conc}. Additional graphs and computational details are included in supporting information (SI). 


\section{Model System} 
\label{sec:preliminfo}

We focus on molecular systems exhibiting nonadiabatic quantum dynamics such as photosynthetic and photovoltaic molecular assemblies, commonly described by the following model Hamiltonian:
\begin{equation}
\hat{H} = \sum_{j=1}^{N_e} \hat{H}_j | j \rangle \langle j| + \sum_{\substack{j,k=1 \\ j \neq k}}^{N_e} \hat{V}_{jk} | j \rangle \langle k|,
\label{eq:genH}
\end{equation}  
Here, $\hat{H}_j = \hat{\bf P}^2/2 + V_j \left( \hat{\bf R} \right)$ is the nuclear Hamiltonian when the system is in diabatic electronic state 
$ | j \rangle$, with index $j$ running over the $N_e$ electronic states ($j=1,2,\ldots,N_e$), while $\hat{\bf R} = \left( \hat{R}_1,..., \hat{R}_{N_n} \right)$ and $\hat{\bf P} = \left( \hat{P}_1,..., \hat{P}_{N_n} \right)$ are the mass-weighted position and momentum operators of the ${N_n} \gg 1$ nuclear DOF, and $\left\{ \hat{V}_{jk} | j \neq k \right\}$ are 
coupling terms between electronic states which can be either nuclear operators (non-Condon case) or constants (Condon case). 
Throughout this paper, a hat over a variable (e.g., $\hat{B}$) indicates an operator quantity and calligraphic font (e.g., ${\cal L}$) indicates a superoperator.


For simplicity, we assume that the initial state of the overall system has the single-product form,
\begin{equation}
\hat{\rho} (0) = \hat{\rho}_n (0) \otimes \hat{\sigma} (0).
\label{init_stat}  
\end{equation}
Here, $\hat{\rho}_n (0) = \text{Tr}_e \{ \hat{\rho} (0)\}$ and $\hat{\sigma}(0) = \text{Tr}_n \{ \hat{\rho} (0)\}$ are the reduced density operators that describe the initial states of nuclear DOF and electronic DOF, respectively, while $\text{Tr}_e \{ \cdot \}$ and $\text{Tr}_n \{ \cdot \}$ represent partial traces over the electronic and nuclear Hilbert spaces, respectively. It should be noted that the methodology presented in this paper is not limited to factorized initial states, as introduced by Eq.~(\ref{init_stat}), and can be applied to arbitrary initial states.\cite{mulvihill19a}

The time-dependent propagation of the initial state, introduced by Eq.~(\ref{init_stat}), according to the Hamiltonian introduced by Eq.~(\ref{eq:genH}) yields the propagated state $\hat{\rho}(t)$ at time $t$ described by the following density operator:
\begin{eqnarray}
\hat{\rho} (t) 
= e^{-i \hat{H} t/\hbar} \hat{\rho}_n (0) \otimes \hat{\sigma} (0) e^{i \hat{H} t/\hbar}
\equiv& e^{-i {\cal L} t/\hbar} \hat{\rho}_n (0) \otimes \hat{\sigma} (0).
\label{over_dyn}  
\end{eqnarray}
Here, ${\cal L} ( \cdot ) = [\hat{H} , \cdot ]$ is the overall  Liouvillian superoperator. The reduced electronic density operator $\hat{\sigma}(t)$ at time $t$ is obtained by tracing out the nuclear, as follows: 
\begin{equation}
\hat{\sigma} (t) = \text{Tr}_n \{ \hat{\rho} (t)\} = \sum_{j,k=1}^{N_e} \sigma_{jk} (t) | j \rangle \langle k |.
\label{reduced_t}
\end{equation}
The electronic populations and coherences are given by $\{ \sigma_{jj} (t) = \langle j | \hat{\sigma} (t) | j \rangle\}$ and $\{ \sigma_{jk} (t) = \langle j | \hat{\sigma} (t) | k \rangle| j \neq k \}$, respectively. These quantities are of particular interest because their time evolution underlies electronic energy, charge, and coherence transfer dynamics, as well as electronic decoherence.


\section{GQMEs in Reduced Dimensionality}
\label{sec:gqme}

The GQME formalism can be applied to derive exact equations of motion for electronic observables while keeping the input regarding other DOF in the system to the minimum necessary to account for their impact on dynamics. 
To this end, we begin with the well-known Nakajima-Zwanzig GQME (whose derivation is outlined in the SI),
\begin{eqnarray}
\frac{d}{dt} \mathcal{P} \hat{\rho} (t) &=& -\frac{i}{\hbar} \mathcal{P} \mathcal{L} \mathcal{P} \hat{\rho} (t) -\frac{1}{\hbar^2} \int_0^t d \tau \mathcal{P} \mathcal{L} e^{-i \mathcal{QL} \tau/\hbar} \mathcal{Q} \mathcal{L} \mathcal{P} \hat{\rho} (t-\tau) -\frac{i}{\hbar} \mathcal{P} \mathcal{L} e^{-i \mathcal{QL} t/\hbar} \mathcal{Q} \hat{\rho}(0),
\label{dPrhodt}
\end{eqnarray}
where ${\cal P}$ is a projection superoperator.~\cite{nakajima58,zwanzig60b}
Here, ${\cal Q} = \mathit{1} - {\cal P}$ is the projection  complementary to ${\cal P}$, $\mathit{1}$ is the identity superoperator, and ${\cal L}$ is the Liouvillian superoperator as in Eq.~(\ref{over_dyn}).
Integrating Eq.~(\ref{dPrhodt}), we obtain the time-dependent projected state ${\cal P}\hat{\rho}(t)$. Importantly, there is a lot of flexibility on the choice of ${\cal P}$ to select the specific quantity of interest.\cite{mulvihill22}

In this paper, we focus on quantities of interest corresponding to a subset of electronic reduced density matrix elements $\left\{  \sigma_{ab} (t) \right\}$ by introducing the following projection operators:
\begin{align}
\mathcal{P}^{\text{set}}\hat{A} &= \sum_{jk\, \in \{ab\}} \Tr\Big\{\Big(|j \rangle\langle k| \otimes \hat{1}_n\Big)^\dagger \hat{A}\Big\} \hat{\rho}_n(0) \otimes |j\rangle\langle k|, \label{eq:Proj_sub}
\end{align}
where $\hat{1}_n$ is the unity operator in the nuclear Hilbert space.  
For example, $\left\{  \sigma_{ab} (t) \right\}$ may include all
$N_e^2$  electronic reduced density matrix elements (i.e., all populations and coherences), 
in which case $\left\{  \sigma_{ab} (t) \right\} \rightarrow \left\{  \sigma_{11} (t), \ldots \sigma_{1N_e} (t), \ldots, \sigma_{N_e 1} (t),\ldots,\sigma_{N_e N_e} (t)\right\}$;
or only the diagonal electronic reduced density matrix elements (i.e., the populations of the corresponding electronic states), in which case $\left\{  \sigma_{ab} (t) \right\} \rightarrow \left\{  \sigma_{11} (t), \ldots ,  \sigma_{N_e N_e} (t)\right\}$ or, just the diagonal term describing the time-dependent population of state $| 1 \rangle$, in which case
$\left\{  \sigma_{ab} (t) \right\}  \rightarrow \left\{  \sigma_{11} (t) \right\}$.

Substituting the projection superoperator $\mathcal{P}^{\text{set}}$
into Eq.~(\ref{dPrhodt}) and tracing over the nuclear and electronic Hilbert spaces, we obtain the following equation of motion for the electronic reduced density matrix elements included in the subset [$\sigma_{jk} (t) \in \left\{ \sigma_{ab} (t) \right\}$]:
\begin{align}
\frac{d}{dt}\sigma_{jk}(t) &= -\frac{i}{\hbar}  \sum_{lm\, \in \{ab\}}\langle{\cal L}_{jk,lm}\rangle_n^0 \,\sigma_{lm}(t)  -\sum_{lm\, \in \{ab\}} \int_0^t d\tau\, {\cal{ K}}_{jk,lm}^{\text{set}}(\tau) \sigma_{lm}(t - \tau) + I_{jk}^{\text{set}}(t).\label{gqme_sub}
\end{align}
Here, $\langle{\cal L}_{jk,lm}\rangle_n^0$, ${\cal K}^{\text{set}}_{jk,lm}(\tau)$, and $I_{jk}^{\text{set}}(t)$ are the matrix elements $(jk,lm)$ of the {\em projected Liouvillian} superoperator, {\em memory kernel} superoperator, and {\em inhomogeneous term} operator, respectively, defined as follows:
\begin{equation}
\langle{\cal L}_{jk,lm}\rangle_n^0 = 
\text{Tr} \Big\{\Big( |j \rangle \langle k | \otimes \hat{1}_n \Big)^\dagger {\cal L} \hat{\rho}_n (0) \otimes | l \rangle \langle m | \Big\},
\label{avL}  
\end{equation}
\begin{eqnarray}
{\cal K}^{\text{set}}_{jk,lm}(\tau) = \frac{1}{\hbar^2} \!\Tr\!\bigg\{\!\Big(|j\rangle\langle k| \otimes \hat{1}_n\!\Big)^\dagger {\cal L}e^{-i{\cal Q}^{\text{set}}{\cal L}\tau/\hbar} {\cal Q}^{\text{set}}{\cal L}\,\hat{\rho}_n(0)\otimes |l\rangle \langle m| \bigg\}, \label{eq:K_sub}
\end{eqnarray}
and
\begin{eqnarray}
I^{\text{set}}_{jk}(t) = -\frac{i}{\hbar} \Tr\bigg\{\!\Big(|j\rangle\langle k| \otimes \hat{1}_n\Big)^\dagger \mathcal{L} e^{-i \mathcal{Q}^{\text{set}}{\cal L} t/\hbar}\Big[\hat{\rho}(0) - \!\!\sum_{lm\, \in \{ab\}}\hat{\rho}_n(0)\otimes |l\rangle\langle m|\, \sigma_{lm}(0)\Big]\!\bigg\}.~~~ \label{eq:I_sub}
\end{eqnarray}

Given that $N_{\text{set}}$ is the number of matrix elements of interest included in $\{\sigma_{ab}(t)\}$ ($1 \le N_{\text{set}} \le N_e^2$), the projected Liouvillian $\langle{\cal L} \rangle_n^0$ and memory kernel ${\cal K}^{\text{set}} (\tau)$ superoperators can be represented by $N_{\text{set}} \times N_{\text{set}}$ matrices, whereas the inhomogeneous term operator $\hat{I}^{\text{set}} (t)$ can be represented by an $N_{\text{set}}$-dimensional vector in Liouville space. 

Calculating the projected Liouvillian is typically straightforward. The memory kernel and the inhomogeneous term satisfy Volterra integral equations, so they can be obtained from the PFIs.\cite{mulvihill22} 
The Volterra equation for the memory kernel is given by 
\begin{equation}
\begin{split}
{\cal K}^{\text{set}}_{jk,lm}(\tau) &= i{\dot {\cal F}}_{jk,lm}(\tau) - \frac{1}{\hbar} \sum_{uv\,\in\{ab\}} {\cal F}_{jk,uv}(\tau) \langle{\cal L}_{uv,lm}\rangle_n^0 
\\ &\qquad +\ i\!\!\sum_{uv\,\in\{ab\}}\int_0^\tau d\tau'\, {\cal F}_{jk,uv}(\tau - \tau'){\cal K}^{\text{set}}_{uv,lm}(\tau'), 
\end{split}\label{eq:Kvolt_sub}
\end{equation}
where the PFIs are given by
\begin{equation}
\begin{split}
{\cal F}_{jk,lm}(\tau) &= \frac{1}{\hbar}\Tr\bigg\{\!\Big(|j\rangle\langle k| \otimes \hat{1}_n\Big)^\dagger {\cal L} e^{-i {\cal L}\tau/\hbar} \hat{\rho}_n(0)\otimes |l\rangle\langle m|\bigg\}, 
\\ {\dot {\cal F}}_{jk,lm}(\tau) &= -\frac{i}{\hbar^2}\Tr\bigg\{\!\Big(|j\rangle\langle k| \otimes \hat{1}_n\Big)^\dagger {\cal L} e^{-i {\cal L}\tau/\hbar} {\cal L} \,\hat{\rho}_n(0)\otimes |l\rangle\langle m|\bigg\}. 
\end{split}\label{eq:FFdotsub}
\end{equation}
The Volterra equation for the inhomogeneous term is given by
\begin{eqnarray}
I^{\text{set}}_{jk}(t) = Z_{jk}(t) + i\sum_{lm\,\in\{ab\}} {\cal F}_{jk,lm}(t) \sigma_{lm}(0) +\ i \sum_{uv\,\in\{ab\}}\int_0^t d\tau\, {\cal F}_{jk,uv}(t - \tau) I^{\text{set}}_{uv}(\tau), \label{eq:Ivolt_sub}
\end{eqnarray}
where the additional PFI $Z_{jk}(t)$ is given by 
\begin{equation}
Z_{jk}(t) = -\frac{i}{\hbar}\Tr\bigg\{\!\Big(|j\rangle\langle k| \otimes \hat{1}_n\Big)^\dagger {\cal L} e^{-i {\cal L}t/\hbar} \hat{\rho}(0)\bigg\}.
\label{eq:Zjk}
\end{equation}
It should be noted that  $Z_{jk}(t) = -i{\cal F}_{jk,\gamma\gamma}(t)$ when the overall initial state is of the commonly encountered form $\hat{\rho}(0) = \hat{\rho}_n(0) \otimes | \gamma \rangle \langle \gamma|$ (where $|\gamma \rangle$ is one of the electronic basis states), as is the case for the applications reported in this paper.
A more detailed discussion of the derivation, properties, and significance of Eqs.~(\ref{eq:Kvolt_sub})-(\ref{eq:Zjk})
can be found in Ref.~\citenum{mulvihill22} and the SI.

Most previous studies have been based on direct calculations of the aforementioned PFIs.\cite{mulvihill19a, mulvihill19b, mulvihill21a, mulvihill21b, mulvihill22,mulvihill19a} However, when using an exact input method, the PFIs can also be accurately obtained as derivatives of the propagator ${\cal U}(\tau) \equiv \Tr_n \left\{ e^{-i{\cal L} \tau /\hbar} \hat{\rho}_n(0) \otimes \hat{1}_e \right\}$ that evolves the electronic reduced density operator, as follows:~\cite{kidon18,mulvihill19a}
\begin{equation}
\hat{\sigma} (\tau) = {\cal U}(\tau) \hat{\sigma} (0),
\end{equation}
with matrix elements,
\begin{equation}
{\cal U}_{jk,lm}(\tau) = \Tr\Big\{\Big(|j\rangle\langle k| \otimes \hat{1}_n\Big)^\dagger e^{-i{\cal L}\tau/\hbar}\hat{\rho}_n(0) \otimes |l \rangle\langle m|\Big\}. \label{eq:U}
\end{equation}
Specifically, we obtain the PFIs $\left\{ {\cal F}_{jk,lm}(\tau), \dot{{\cal F}}_{jk,lm}(\tau) \right\}$ from $\left\{ {\cal U}_{jk,lm}(\tau) \right\}$, as follows:\cite{kidon18,mulvihill19a}
\begin{equation}
{\cal F}_{jk,lm}(\tau) = i\dot{\cal U}_{jk,lm}(\tau), \qquad \dot{\cal F}_{jk,lm}(\tau) = i\ddot{\cal U}_{jk,lm}(\tau). \label{eq:FFdotU}
\end{equation}

PFIs $\left\{ {\cal F}_{jk,lm}(\tau), \dot{\cal F}_{jk,lm}(\tau) \right\}$ obtained from ${\cal U}(\tau)$ generate exact memory kernels and inhomogeneous terms when ${\cal U}(\tau)$ is obtained from exact inputs. Therefore, we obtain them in terms of numerical derivatives of ${\cal U}(\tau)$ obtained from TT-TFD simulations, as described in Sec.~\ref{sec:tt-tfd}.



\section{Tensor-Train Thermo-Field Dynamics}
\label{sec:tt-tfd}

\subsection{Hamiltonian}

The molecular Hamiltonian introduced by Eq.~(\ref{eq:genH}) can also be written as a sum of a purely electronic Hamiltonian $\hat{H}_e \otimes \hat{1}_n$ plus a purely nuclear Hamiltonian $\hat{1}_e \otimes \hat{H}_n$ and an interaction term between the electronic and nuclear DOF, $\hat{H}_{en}$:
\begin{equation}
\begin{split}
\hat{H}&=\hat{H}_e \otimes \hat{1}_n + \hat{1}_e \otimes \hat{H}_{n}+\hat{H}_{en}.
\end{split}
\label{H_en}
\end{equation}
It should be noted that this division is not unique, in the sense that different choices of $\hat{H}_e$, $\hat{H}_n$, and $\hat{H}_{en}$ are possible.\cite{mulvihill19a} However, the results are invariant to those choices when a quantum-mechanically exact method like TT-TFD is applied since no physical or {\em ad hoc} approximation is introduced.

\subsection{Thermo-field dynamics method}
We start out by noting that the dynamics of $\hat{\rho}(t)$ governed by the Hamiltonian of the form of Eq.~(\ref{H_en}) is described by the quantum Liouville equation,
\begin{equation}\label{eq:Liou}
\frac{d}{dt}\hat{\rho}(t)=-\frac{i}{\hbar}[\hat{H},\hat{\rho}(t)].
\end{equation}
The TT-TFD method \cite{Gelin2017,Borrelli2016,Borrelli2017,Borrelli2021} provides a general, numerically exact approach to solve Eq.~\eqref{eq:Liou} that is particularly efficient when  $\hat{\rho}(t)$ can be represented as a low rank matrix product state. In our simulations, the state is described by
$\hat{\rho}^{1/2} (t)$ (instead of $\hat{\rho} (t)$), represented as a tensor-train vector in an extended Hilbert space (the so-called double Hilbert space described below). The Liouville equation given in Eq.~(\ref{eq:Liou}) is replaced by an equivalent equation of motion for $\hat{\rho}^{1/2} (t)$, which can be written in the form of a Schr\"{o}dinger-like equation in the double Hilbert space. 
For a high-dimensional system, computational efficiency is achieved by using a tensor-train representation~\cite{Oseledets2011,Oseledets2010,Grasedyck2009,Hackbusch2009,greene17,Lyu2022,soley2021functional,soley2021iterative}
of the extended state vector $\hat{\rho}^{1/2} (t)$. The remainder of this section outlines the TT-TFD methodology used for calculating the PFIs needed to obtain the memory kernel and inhomogeneous term of the GQMEs. 

%
The initial density operator of the overall system is of the form introduced by Eq.~\eqref{init_stat}. The initial electronic density operator is given by $\hat{\sigma}(0)=|\gamma\rangle\langle\gamma|$, where $| \gamma \rangle$ is one of the electronic basis states, while the initial nuclear density operator is 
$\hat{\rho}_n(0)=e^{-\beta \hat{H}_n}/Z_n(\beta)$, where $Z_n(\beta)=\text{Tr}_n\{e^{-\beta H_n}\}$. Therefore,
\begin{equation}\label{eq:rho0}
\hat{\rho}(0)
=|\gamma\rangle\langle\gamma|\otimes 
\frac{e^{-\beta {\hat{H}_n}}}{Z_n(\beta)}.
\end{equation}
We note, however, that the TT-TFD method is not restricted to initial states of this simple form and can be analogously applied to propagate any arbitrary initial state. 

The TFD representation is only applied to the nuclear density operator of the system since the same dynamics is obtained for the initial state introduced by Eq.~\eqref{eq:rho0} regardless of whether the electronic density operator is included or not in the TFD representation.~\cite{Borrelli2016} We let $\{|k \rangle\}$ be an orthonormal basis that spans the {\em physical} nuclear Hilbert space $\mathscr{H}_n$ and $\{|\tilde{k} \rangle\}$ an orthonormal basis that spans a {\em fictitious} nuclear Hilbert space (also known as the tilde space) $\tilde{\mathscr{H}}_n$, which is an exact replica of $\mathscr{H}_n$.
Next, we define the so-called {\em nuclear thermal vacuum state}:
\begin{equation}
|0_n(\beta)\rangle=\frac{e^{-\beta \hat{H}_n/2}}{\sqrt{Z(\beta)}}\sum_{\tilde{k}=k}
|k\rangle \otimes |\tilde{k}\rangle,
\label{eq:nts}
\end{equation}
where it should be noted that the sum includes only terms  $|k\rangle \otimes |\tilde{k}\rangle$ with $\tilde{k}=k$, so that $\sum_{\substack{\tilde{k}=k}} |k\rangle \otimes |\tilde{k}\rangle = |0\rangle \otimes |\tilde{0}\rangle + |1\rangle \otimes |\tilde{1}\rangle +...\;$. 
We note that $\hat{\rho}_n(0)$ can be obtained from $|0_n(\beta)\rangle$, upon taking the outer product with its dual and tracing out the fictional degrees of freedom as follows:
\begin{equation}\label{eq:0nbeta}
\text{Tr}_f\Big\{|0_n(\beta)\rangle\langle 0_n(\beta)|\Big\}
=
\hat{\rho}_n(0),
\end{equation}
where $\text{Tr}_f\{\cdot\}$ is the partial trace over states $|\tilde{k}\rangle$ in the tilde space $\tilde{\mathscr{H}}_n$.

Substituting Eq.~\eqref{eq:0nbeta} into Eq.~\eqref{eq:rho0}, we obtain the initial density operator of the overall system $\hat{\rho}(0)$ represented in terms of the ket vector $|\psi_\gamma(\beta,0)\rangle\equiv|\gamma\rangle \otimes |0_n(\beta)\rangle$, as follows:
\begin{equation}\label{eq:TFDrho0}
\hat{\rho}(0)
=
\text{Tr}_f\Big\{|\psi_\gamma(\beta,0)\rangle\langle\psi_\gamma(\beta,0)|\Big\}. 
\end{equation}
Note that in Eq.~\eqref{eq:TFDrho0}, only the initially thermalized nuclear density operator is represented by a ket vector in the double space
$\mathscr{H}_n \otimes \tilde{\mathscr{H}}_n$;
whereas the initial electronic density operator $|\gamma\rangle\langle\gamma|$ corresponds to a pure state in the electronic Hilbert space. 

We define the overall system ket vector $|\psi_\gamma(\beta,t)\rangle$ such that
\begin{equation}
\label{eq:TFDrhot}
\hat{\rho}(t)
=
\text{Tr}_f\Big\{|\psi_\gamma(\beta,t)\rangle\langle\psi_\gamma(\beta,t)|\Big\},
\end{equation}
where $\hat{\rho}(t)$ evolves according to the Liouville equation Eq.~\eqref{eq:Liou}. This can be fulfilled by evolving $|\psi_\gamma(\beta,t)\rangle$ 
according to the so-called TFD Schr\"{o}dinger equation (as shown in the SI),
\begin{equation}\label{eq:TFDpsi}
\frac{d}{dt}|\psi_\gamma(\beta,t)\rangle=-\frac{i}{\hbar}\bar{H}|\psi_\gamma(\beta,t)\rangle,
\end{equation}
where $\bar{H}=\hat{H}\otimes\tilde{1}_n$, with $\tilde{1}_n=\sum_{\tilde{k}}|\tilde{k}\rangle\langle\tilde{k}|$ the identity operator of the tilde space. 
%
%
Moreover, we note that the same physical system dynamics can be obtained by defining $\bar{H}$ in Eq.~\eqref{eq:TFDpsi}, as follows:
\begin{equation}
\bar{H}=\hat{H}\otimes\tilde{1}_n-\hat{1}\otimes\tilde{H}_n,
\end{equation}
where $\hat{1} = \hat{1}_n \otimes \hat{1}_e$. Remarkably, $\tilde{H}_n$ can be {\em any} operator in the nuclear tilde space since $\tilde{H}_n$
does not impact kets in the physical space and its effect on the dynamics vanishes upon taking the partial trace over states in the tilde space.~\cite{Borrelli2016} 

The preparation of the initial thermal wavepacket $|\psi_{\gamma}(\beta,0)\rangle$, according to Eqs.~\eqref{eq:0nbeta} and~\eqref{eq:nts}, requires the explicit evaluation of the quantum Boltzmann operator, which can be computationally challenging for systems with high dimensionality. However, when the initial nuclear Hamiltonian is harmonic, the initial thermal wavepacket can be obtained by taking advantage of the {\em thermal Bogoliubov transformation}. Therefore, we can generate the nuclear thermal vacuum state from the double space ground state $|0_n,\tilde{0}_n\rangle$ using the following unitary transformation,
\begin{equation}\label{eq:Ther_Bog1}
|\psi_{\gamma}(\beta,0)\rangle=|\gamma\rangle \otimes e^{-i\hat{G}}|0_n,\tilde{0}_n\rangle,
\end{equation}
where $\hat{G}$ is given by:\cite{Takahashi1996,Gelin2017,Borrelli2021}
\begin{equation}\label{eq:rot_G1}
\hat{G}=-i\sum_j\theta_j(\hat a_j\tilde{a}_j - \hat a_j^\dagger \tilde{a}_j^\dagger),
\end{equation}
with $\theta_j=\text{arctanh} \left( e^{-\beta\omega_j/2} \right)$, where $\{ \hat{a}_j , \hat{a}_j^\dagger\}$
and $\{ \tilde{a}_j , \tilde{a}_j^\dagger\}$ are
the creation and annihilation operators associated with the $j$-th nuclear DOF in the physical and tilde Hilbert spaces, respectively. 

Substituting Eqs.~\eqref{eq:Ther_Bog1} and \eqref{eq:rot_G1} into Eq.~\eqref{eq:TFDpsi} we obtain:
\begin{equation}\label{eq:theta_TFD1}
\frac{d}{dt}|\psi_{\theta,\gamma}(\beta,t)\rangle
=-\frac{i}{\hbar}
\bar{H}_\theta|\psi_{\theta,\gamma}(\beta,t)\rangle
,
\end{equation}
with $|\psi_{\theta,\gamma}(\beta,0)\rangle=|\gamma\rangle \otimes |0,\tilde{0}\rangle$, $|\psi_{\theta,\gamma}(\beta, t)\rangle=e^{i\hat{G}}|\psi_{\gamma}(\beta,t)\rangle$, and $\bar{H}_\theta$ is defined as:
\begin{equation}\label{eq:hatHtheta}
\bar{H}_\theta=e^{i\hat{G}}\bar{H}e^{-i\hat{G}}.
\end{equation}

The time-dependent thermal state $|\psi_{\theta,\gamma}(\beta,t)\rangle$ is represented as:
\begin{equation}
|\psi_{\theta,\gamma}(\beta,t)\rangle=\sum_{j_1,...,j_d}^{n_1,...,n_d}X(\beta,t ; j_1,...,j_d)|j_1\rangle \otimes... \otimes |j_d\rangle,
\end{equation}
where $d=1+2N_n$ is the overall number of DOF and $\{|j_k\rangle\}$ is the basis set with $k=1,...,d$. We determined the size of the basis according to the convergence test, including two electronic state eigenvectors and the 10 nuclear harmonic eigenvectors for the nuclear DOF. 

The time- and temperature-dependent expansion coefficients $\{ X(\beta,t;j_1,...,j_d) \}$ correspond to an $n_1\times...\times n_d$ complex array which requires storage space and computational effort that grows exponentially with $d$. Thus, we avoid the curse of dimensionality by implementing the TFD wavepacket in the tensor-train (TT) format.\cite{Oseledets2011,Oseledets2010,Grasedyck2009,Hackbusch2009,greene17,Lyu2022,soley2021functional,soley2021iterative}

\subsection{TT Format}

The TT format of $X\in\mathbb{C}^{n_1\times...\times n_d}$ involves a train-like product of $d$ tensor cores which are 3-mode tensors $X_i\in\mathbb{C}^{r_{i-1}\times n_i\times r_i}$, with $r_0=r_d=1$. Any particular element $X(j_1,...,j_d)$ can be evaluated by multiplication of the cores, as follows:
\begin{equation}
X(j_1,...,j_d)=\sum_{a_0=1}^{r_0}\sum_{a_1=1}^{r_1}...\sum_{a_d=1}^{r_d}X_1(a_0,j_1,a_1)X_2(a_1,j_2,a_2)...X_d(a_{d-1},j_d,a_d).
\label{ttformat}
\end{equation}
This can also be written in compact matrix product notation, as follows:
\begin{equation}
X(j_1,...,j_d)=\mathbf{X}_1(j_1)\mathbf{X}_2(j_2)...\mathbf{X}_d(j_d),
\end{equation}
with matrix $\mathbf{X}_i(j_i)\in\mathbb{C}^{r_{i-1}\times r_i}$ defining the $j_i^{th}$ slice of $X_i$. 

The central idea of the TT format is to generalize the concept of factorization. Each physical dimension $i$ is factorized as an individual core (i.e., $X_i$). Entanglement with other physical dimensions is established through the auxiliary indices $a_{i-1}$ and $a_i$. The TT-ranks $r_0,...,r_d$
introduced by Eq.~(\ref{ttformat}) remain small for a low level of entanglement and when they are $r_0=...=r_d=1$, the TT format of $X$ is a factorizable product. 

Eq.~(\ref{ttformat}) shows that the TT format allows for compressed representations of $X$ since it requires storage of $X_1,...,X_d$, with $dn\tilde{r}^2$ elements when $r_1=...=r_{d-1}=\tilde{r}$ and $n_1=...=n_d=n$. For small $\tilde{r}$, such a representation bypasses the need to explicitly store all $n^d$ elements of $X$, thus offering an exponential advantage in storage and computational effort. 

In TT-TFD, the initial state $|\psi_{\theta,\gamma}(\beta,0)\rangle=|\gamma\rangle \otimes |0,\tilde{0}\rangle$ takes an initial single-product form and is prepared as a rank-1 tensor train. The transformed TFD Schr\"{o}dinger equation is then solved with the TT-KSL method.\cite{Lubich2015,Lubich2014} The TT-KSL propagator evolves the wavepacket according to the time-dependent variational principle (TDVP) by evolving the time-dependent state on a fixed-rank TT manifold. Comparisons to other TT propagators have shown that TT-KSL is quite accurate and efficient.\cite{Li2020,Lyu2022}

\subsection{Projection-free inputs from TT-TFD}

The PFIs required for calculating the memory kernel and inhomogeneous term of the GQME are computed by using the TT-TFD methodology.
According to Eq.~\eqref{eq:U}, the matrix elements ${\cal U}_{jk,lm}(\tau)$ are obtained, as follows:
\begin{equation}
\begin{split}
\mathcal{U}_{jk,lm}(\tau)&=\text{Tr}_{e,n}\Big\{e^{-i\hat{H}\tau/\hbar}\hat{\rho}_n(0)|l\rangle\langle m|e^{i\hat{H}\tau/\hbar}(|k\rangle\langle j|\otimes \hat{1}_n)\Big\}.
\end{split}
\end{equation}
Since TT-TFD requires an initial electronic state that is in a pure state $|\gamma\rangle$, in the following we write $|l\rangle\langle m|$ as $|\gamma\rangle\langle \gamma|$; however, we note that all ${\cal U}(\tau)$ elements with off-diagonal initial electronic density matrices can be expressed as linear combinations of pure-state populations (see the SI).   

With Eqs.~\eqref{eq:Ther_Bog1} and \eqref{eq:theta_TFD1}, we use $|k\rangle\langle j|j\rangle\langle j|=|k\rangle\langle j|$ to rewrite ${\cal U}_{jk,\gamma\gamma}(\tau)$ as
\begin{equation}\label{U_TFD}
\begin{split}
\mathcal{U}_{jk,\gamma\gamma}(\tau)&=\text{Tr}_{e,n}\Big\{e^{-i\hat{H}\tau/\hbar}\hat{\rho}_n(0)|\gamma\rangle\langle \gamma|e^{i\hat{H}\tau/\hbar}(|k\rangle\langle j|\otimes \hat{1}_n)\Big\}\\
&=\text{Tr}_{e,n}\Big\{e^{-i\hat{H}\tau/\hbar}\hat{\rho}_n(0)|\gamma\rangle\langle \gamma|e^{i\hat{H}\tau/\hbar}(|k\rangle\langle j|\otimes \hat{1}_n)(|j\rangle\langle j|\otimes \hat{1}_n)\Big\}.
\end{split}
\end{equation}
From this equation, noting that $\Tr_f\Big\{|\psi_\gamma(\beta,\tau)\rangle\langle\psi_\gamma(\beta,\tau)|\Big\} = e^{-i\hat{H}\tau/\hbar}\hat{\rho}_n(0)|\gamma\rangle\langle \gamma|e^{i\hat{H}\tau/\hbar}$, we perform a cyclic permutation to obtain
\begin{equation}
{\cal U}_{jk,\gamma\gamma}(\tau) =\text{Tr}_{e,n}\bigg\{\text{Tr}_f\Big\{(|j\rangle\langle j|\otimes \hat{1}_n)|\psi_{\gamma}(\beta, t)\rangle\langle\psi_{\gamma}(\beta,\tau)|(|k\rangle\langle j|\otimes \hat{1}_n)\Big\}\bigg\}.
\end{equation}
From here, we use $|\psi_{jk,\gamma\gamma}(\tau)\rangle=(|k\rangle\langle j|\otimes \hat{1}_n)|\psi_\gamma(\beta,\tau)\rangle$ and $|\psi_{\theta,jk\gamma\gamma}(\tau)\rangle=e^{iG}|\psi_{jk,\gamma\gamma}(\tau)\rangle=(|k\rangle\langle j|\otimes \hat{1}_n)|\psi_{\gamma,\theta}(\beta,\tau)\rangle$ to obtain,
\begin{equation}
\begin{split}
\mathcal{U}_{jk,\gamma\gamma}(\tau)&=\text{Tr}_{e,n,f}\Big\{|\psi_{jj,\gamma\gamma}(\beta, \tau)\rangle\langle\psi_{jk,\gamma\gamma}(\beta,\tau)|\Big\},\\
&=\langle\psi_{jj,\gamma\gamma}(\tau)|\psi_{jk,\gamma\gamma}(\tau)\rangle,\\
&=\langle\psi_{\theta,jj\gamma\gamma}(\tau)|\psi_{\theta,jk\gamma\gamma}(\tau)\rangle,
\end{split}
\end{equation}
which provides the elements of $\mathcal{U}(\tau)$ after obtaining $|\psi_{\gamma,\theta}(\beta,\tau)\rangle$ by integrating Eq.~\eqref{eq:theta_TFD1}.


\section{Applications}
\label{sec:spin-boson}

In this section, we report simulations of electronic population dynamics based on four types of GQMEs. The equations correspond to different subsets of electronic reduced density matrix elements used to describe the underlying dynamics (see Sec.~\ref{sec:gqme}).
As described in Sec.~\ref{sec:tt-tfd}, the memory kernels and inhomogeneous terms are calculated from PFIs obtained via the quantum-mechanically exact TT-TFD method (see Sec.~\ref{sec:tt-tfd}) as applied to five different realizations of a benchmark spin-boson model Hamiltonian. We also compare the quantum-mechanically exact memory kernels and inhomogeneous terms obtained with TT-TFD inputs to calculations based on an approximate linearized semiclassical (LSC) method.\cite{mulvihill22} 



The reduced electronic density matrix for the spin-boson model, introduced in Sec.~\ref{subsec:spinboson}, consists of four matrix elements, $\{ \sigma_{DD}, \sigma_{DA}, \sigma_{AD}, \sigma_{AA} \}$, where $| D \rangle$ and $| A \rangle$ correspond to the {\em donor} and {\em acceptor} electronic states, respectively. We consider GQMEs for the following four subsets of matrix elements:
(1) $\{ \sigma_{DD}, \sigma_{DA}, \sigma_{AD},  \sigma_{AA}\}$ (the full density matrix);
(2) $\{ \sigma_{DD}, \sigma_{AA} \}$ (the populations-only subset);
(3) $\{ \sigma_{DD} \}$ (the donor single-population subset); and
(4) $\{ \sigma_{AA} \}$ (the acceptor single-population subset). 
The TT-TFD-based PFIs, obtained by taking numerical derivatives of the time evolution operator ${\cal U}(\tau)$ [see Eq.~\eqref{eq:U}], are compared to PFIs obtained via an LSC-based method denoted LSCII [sometimes also referred to as the LSC initial value representation (LSC-IVR) method\cite{sun98}].
 Ref.~\citenum{mulvihill19b} provides a detailed discussion of the protocols used for calculating PFIs via LSCII.


\subsection{Spin-Boson Models}
\label{subsec:spinboson}

The spin-boson model provides a useful framework for studying molecular systems where the dynamics involves two coupled electronic states. In the simplest form, the electronic coupling is independent of the nuclear coordinates (the so-called Condon approximation). The nuclear motion in each electronic state is described by harmonic potential energy surfaces (PESs) with distinct equilibrium energies and equilibrium positions. As such, the spin-boson model has been widely used for describing a wide range of chemical dynamical processes, including charge and energy transfer (e.g., Marcus theory), nonadiabatic dynamics, photochemistry, spin energy relaxation and dephasing, vibrational energy relaxation, and, more recently, polaritonic chemistry where the photonic DOF can be described as harmonic oscillators and therefore grouped with the nuclear DOF.\cite{leggett87, breuer02, nitzan06, weiss12, saller22} 


The spin-boson Hamiltonian is defined according to Eq.~\eqref{eq:genH} with $\{ \hat{H}_j \}$ and $\{ \hat{V}_{jk} \rightarrow V_{jk} \}$ defined, as follows:
\begin{equation}
\begin{split}
\hat{H}_1 &\equiv \hat{H}_D = \epsilon + \sum_{k = 1}^{N_n} \frac{\hat{P}_k^2}{2} + \frac{1}{2}\omega_k^2\hat{R}_k^2 -c_k \hat{R}_k,
\\\hat{H}_2 &\equiv \hat{H}_A = -\epsilon + \sum_{k = 1}^{N_n} \frac{\hat{P}_k^2}{2} + \frac{1}{2}\omega_k^2\hat{R}_k^2 +c_k \hat{R}_k,
\\ 
V_{12} &\equiv V_{DA} = V_{21} \equiv V_{AD} = \Gamma.
\end{split} 
\label{eq:SBham}
\end{equation}  
Here, $2 \epsilon$ is the energy difference between the donor ($D$) and acceptor ($A$) states with nuclear coordinates at equilibrium, and the electronic coupling between donor and acceptor states is defined by the positive constant $\Gamma$ (Condon approximation).

The frequencies $\{\omega_k\}$ and electron-phonon coupling coefficients, $\{c_k\}$ of the nuclear modes are sampled from an Ohmic spectral density with an exponential cutoff:
\begin{align}
  J (\omega) &= \frac{\pi}{2} \sum_{k=1}^{N_n} \frac{c_k^2}{\omega_k} \delta(\omega-\omega_k) ~  \stackrel{\raisebox{1pt} {\text{\footnotesize$N_n \rightarrow \infty$}}}{\xrightarrow{\hspace*{0.75cm}}} ~ \frac{\pi\hbar}{2}
 \xi \omega e^{-\omega/\omega_c}.
 \label{ohmic}
\end{align}
Here, $\xi$ is the Kondo parameter, which determines the electron-phonon coupling strength, and $\omega_c$ is the cutoff frequency which determines the characteristic vibrational frequency. A discrete set of $N_n$ nuclear mode frequencies, $\{\omega_k\}$, and coupling coefficients, $\{c_k\}$, are sampled from the spectral density, introduced by Eq.~(\ref{ohmic}).~\cite{mulvihill19a}. 
The Hamiltonian introduced by Eqs.~\eqref{eq:genH} and~\eqref{eq:SBham} can be rewritten in terms of the harmonic oscillator raising and lowering operators, as follows:
\begin{equation}
\hat{H}=\epsilon\hat{\sigma}_z+\Gamma\hat{\sigma}_x+\sum_{k=1}^{N_n}\omega_k \hat{a}_k^\dagger \hat{a}_k-\sigma_z\frac{c_k}{\sqrt{2\omega_k}}(\hat{a}_k+\hat{a}_k^\dagger).
\end{equation}
The corresponding rotated double space Hamiltonian $\bar{H}_\theta$ introduced by Eq.~\eqref{eq:hatHtheta} can then be obtained in closed 
form, as follows:\cite{Borrelli2021,Ren2022}
\begin{equation}\label{eq:SBthetaH}
\bar{H}_\theta=\epsilon\hat{\sigma}_z+\Gamma\hat{\sigma}_x+\sum_{k=1}^{N_n}\omega_k(\hat{a}_k^\dagger \hat{a}_k-\tilde{a}_k^\dagger\tilde{a}_k)-\frac{\sigma_z c_k}{\sqrt{2\omega_k}}\left((\hat{a}_k+\hat{a}_k^\dagger)\text{cosh}(\theta_k)+(\tilde{a}_k+\tilde{a}_k^\dagger)\text{sinh}(\theta_k) \right),
\end{equation}
where $\hat{\sigma}_x$ and $\hat{\sigma}_z$ are the $x$- and $z$-Pauli matrices. Using Eq.~\eqref{eq:SBthetaH} in place of the mathematically equivalent Eq.~\eqref{eq:hatHtheta} facilitates the implementation of TT-TFD by avoiding the need to calculate $e^{i\hat{G}}$ and $e^{-i\hat{G}}$ numerically.

The initial state is defined according to Eq.~(\ref{init_stat}) with the initial electronic state $\hat{\sigma} (0) = | D \rangle \langle D |$ and the initial nuclear state:
\begin{equation}
  \hat{\rho}_n (0) 
  =\frac{e^{-\beta (\hat{H}_D + \hat{H}_A)/2}}{\text{Tr}_n \Big\{ e^{-\beta(\hat{H}_D + \hat{H}_A)/2} \Big\}}.
  \label{init-nuc}
\end{equation}

Five different models are analyzed, as defined by the sets of parameters listed in Table \ref{tab:parameters}, corresponding to models 1, 2, 3, 4, and 6 of Refs.~\citenum{mulvihill19a}, \citenum{mulvihill19b}, and \citenum{mulvihill22}. Model 5 was not included because the reference exact results are known only for short final times compared to the lifetime of the electronic relaxation dynamics. 
Models 1-3 correspond to systems with a finite energy bias between the donor and acceptor states ($\epsilon = 1.0$), differing with respect to the value of $\omega_c$.
Model 4 corresponds to a biased system ($\epsilon = 1.0$) with a higher Kondo parameter ($\xi=0.4$) relative to models 1-3 ($\xi=0.1$). Model 6 corresponds to a unbiased system ($\epsilon = 0.0$). All results are obtained using an integration time step $\Delta t = 1.50083 \times 10^{-3}\,\Gamma^{-1}$. 
Quantum-mechanically exact QuAPI results for models 1-4 are from Ref.~\citenum{kelly15}, and for model 6 from Ref.~\citenum{kelly13}. 

\begin{table}
\centering
\caption{{\bf Spin-Boson Model and Simulation Parameters}}
\def\arraystretch{1.125}
\begin{tabular}{|c||c|c|c|c|c||c|c|c|}
  \hline
& \multicolumn{5}{c||}{Model Parameters} & \multicolumn{3}{c|}{Numerical Parameters}
\\ \hline Model $\#$ & $\epsilon$ & $\Gamma$ & $\beta$ & $\xi$ & $\omega_c$  &\ $\omega_{\text{max}}$\ \ & \ \ $N_n$\ \ & $\Delta t$  
  \\ \hline
  1  & 1.0  & 1.0  & 5.0  & 0.1  & 1.0  & 5   & 60 & 1.50083 $\times 10^{-3}$
  \\ \hline
  2  & 1.0  & 1.0  & 5.0  & 0.1  & 2.0  & 10  & 60 & 1.50083 $\times 10^{-3}$
  \\ \hline
  3  & 1.0  & 1.0  & 5.0  & 0.1  & 7.5  & 36  & 60 & 1.50083 $\times 10^{-3}$
  \\ \hline
  4  & 1.0  & 1.0  & 5.0  & 0.4  & 2.0  & 10  & 60 & 1.50083 $\times 10^{-3}$
  \\ \hline
  6  & 0.0  & 1.0  & 5.0  & 0.2  & 2.5  & 12  & 60 & 1.50083 $\times 10^{-3}$
  \\ \hline
\end{tabular} 
\label{tab:parameters}
\end{table}


\subsection{GQMEs}
\label{subsec:gqme}

The following subsections outline four types of GQMEs examined by our simulations, corresponding to the analysis of quantum dynamics for different subsets of electronic reduced density matrix elements.  

\subsubsection{Full Set: GQME for All Electronic Density Matrix Elements}

Here, we consider the GQME when the quantities of interest include all four reduced electronic density matrix elements, 
$\{\sigma_{ab}(t)\} = \{\sigma_{DD}(t), \sigma_{DA}(t), \sigma_{AD}(t), \sigma_{AA}(t)\}$:
\begin{equation}
\frac{d}{dt} \sigma_{jk} (t) 
=
-\frac{i}{\hbar} 
\sum_{l,m = 1}^{N_e = 2} 
\langle \mathcal{L}_{jk,lm} \rangle_n^0  \sigma_{lm} (t) 
- 
\sum_{l,m = 1}^{N_e = 2} \int_0^t d\tau\ \mathcal{K}^{\text{full}}_{jk,lm}(\tau) \sigma_{lm} (t - \tau),
\label{eq:mGQME}
\end{equation}
where $jk \in \{ DD,DA,AD,AA \}$.
The memory kernel superoperator $\mathcal{K}^{\text{full}} (\tau)$ is represented 
by an $N_e^2 \times N_e^2 = 4 \times 4$ time-dependent matrix 
whose matrix elements are obtained by solving the following Volterra equation:
\begin{equation}\label{eq:K_full}
\mathcal{K}^{\text{full}}_{jk,lm} (\tau) 
=
i \dot{\mathcal{F}}_{jk,lm} (\tau) -\frac{1}{\hbar} \sum_{u,v = 1}^{N_e = 2} \mathcal{F}_{jk,uv} (\tau) \langle \mathcal{L}_{uv,lm} \rangle_n^0
+ i \sum_{u,v = 1}^{N_e = 2} \int_0^\tau d\tau' \mathcal{F}_{jk,uv} (\tau - \tau') {\cal K}^{\text{full}}_{uv,lm} (\tau'), 
\end{equation}
where the PFIs $\{ \mathcal{F}_{jk,lm} (\tau) \}$ and $\{ \dot{\mathcal{F}}_{jk,lm} (\tau) \}$ are introduced by Eq.~\eqref{eq:FFdotsub}. 


\subsubsection{Populations-Only: GQME for Diagonal Elements of the Reduced Electronic Density Matrix}

Here, we consider the GQME for the quantities of interest that includes only the diagonal matrix elements of the reduced electronic density matrix (i.e., the populations-only GQME), such that $\{\sigma_{ab}(t)\} = \{\sigma_{DD}(t), \sigma_{AA}(t)\}$:
\begin{eqnarray}
\frac{d}{dt}{\sigma}_{jj}(t) = 
-\sum_{k = 1}^{N_e = 2} 
\int_0^t d\tau\, \mathcal{K}_{jj,kk}^{\text{pop}}(\tau) \sigma_{kk}(t - \tau), 
\label{gqme_pop}
\end{eqnarray}
where $j \in \{ D,A\} $.
The memory kernel superoperator $\mathcal{K}^{\text{pop}}(\tau)$ is represented by an $N_e \times N_e = 2 \times 2$ time-dependent matrix, with individual matrix elements obtained by solving the following Volterra equation: 
\begin{eqnarray}
\mathcal{K}^{\text{pop}}_{jj,kk}(\tau) = i\dot{ \mathcal{F}}_{jj,kk}(\tau) +\ i\sum_{\lambda = 1}^{N_e}\int_0^\tau d\tau'\, \mathcal{F}_{jj,\lambda\lambda}(\tau - \tau') \mathcal{K}^{\text{pop}}_{\lambda\lambda,kk}(\tau'), \label{eq:Kvolt_pop}
\end{eqnarray}
where the PFIs $\{ \mathcal{F}_{jj,kk}(\tau) \}$ and $\{ \dot{ \mathcal{F}}_{jj,kk}(\tau) \}$ 
are introduced by Eq.~(\ref{eq:FFdotsub}). 


\subsubsection{Single-Population Scalar: GQMEs for One Diagonal Element of the Reduced Electronic Density Matrix}

Finally, we consider the two single-population scalar GQMEs for the case where the subset includes either only the population of the donor state ($\sigma_{DD}$)
or only the population of the acceptor state ($\sigma_{AA}$), such that $\{\sigma_{ab}(t)\} = \{\sigma_{DD}(t) \}$ or 
$\{\sigma_{ab}(t)\} = \{\sigma_{AA}(t) \}$, respectively:
\begin{eqnarray}
\frac{d}{dt} \sigma_{DD} (t) &=& -\int_0^t d\tau \mathcal{K}^{\text{donor}}_{DD,DD} (\tau) \sigma_{DD} (t-\tau),
\label{gqme_DD}    
\\ \frac{d}{dt} \sigma_{AA} (t) &=& -\int_0^t d\tau \mathcal{K}^{\text{acceptor}}_{AA,AA}  (\tau) \sigma_{AA} (t-\tau) + I^{\text{acceptor}}_{AA} (t).
\label{gqme_AA}    
\end{eqnarray}  
It should be noted that the inhomogeneous term does not vanish in the case where $\{\sigma_{ab}(t)\} = \{\sigma_{AA}(t) \}$. It should also be noted that the memory kernels $\mathcal{K}^{\text{donor}}_{DD,DD}  (\tau)$
and $\mathcal{K}^{\text{acceptor}}_{AA,AA}  (\tau)$, as well the inhomogeneous term $I^{\text{acceptor}}_{AA} (t)$, are scalar in this case and can be obtained by solving the following Volterra equations:
\begin{align}
\mathcal{K}^{\text{donor}}_{DD,DD} (\tau) &= i\dot{\mathcal{F}}_{DD,DD}(\tau) + i\int_0^\tau d\tau'\, \mathcal{F}_{DD,DD}(\tau - \tau') \mathcal{K}^{\text{donor}}_{DD,DD} (\tau'),
\label{eq:Kvolt_DD}
\\ \mathcal{K}^{\text{acceptor}}_{AA,AA}(\tau) &= i\dot{ \mathcal{F}}_{AA,AA}(\tau) + i\int_0^\tau d\tau'\, \mathcal{F}_{AA,AA}(\tau - \tau') \mathcal{K}^{\text{acceptor}}_{AA,AA}(\tau'),
\label{eq:Kvolt_AA}
\\ I^{\text{acceptor}}_{AA}(t) &= -i\mathcal{ F}_{AA,DD}(t) + i \int_0^t d\tau\, \mathcal{F}_{AA,AA}(t - \tau) I^{\text{acceptor}}_{AA}(\tau),
\label{eq:Ivolt_AA}
\end{align}
where the PFIs $\mathcal{F}_{DD,DD}$, $\mathcal{F}_{AA,AA}$, $\dot {\mathcal{F}}_{DD,DD}$, $\dot{\mathcal{F}}_{DD,DD}$, and $\mathcal{F}_{AA,DD}(\tau)$ are defined by Eq.~\eqref{eq:FFdotsub}.

\subsection{Input Methods}

It is important to note that the four types of GQMEs, outlined in the previous subsections, call for the same input of PFIs defined by Eq.~\eqref{eq:FFdotsub}. The different types of GQMEs differ only with respect to the specific matrix elements of $\mathcal{F} (\tau)$ and $\dot{\mathcal{F}} (\tau)$ that are required to calculate the memory kernel and inhomogeneous term. 
For example, calculating the memory kernel for evolving the full set of reduced density matrix elements according to Eq.~(\ref{eq:mGQME}) requires calculating all 16 matrix elements of $\mathcal{F} (\tau)$ and $\dot{\mathcal{F}} (\tau)$.
In contrast, calculating the memory kernel of the donor single-population GQME, Eq.~(\ref{gqme_DD}), requires only a single matrix element of each of the matrices representing $\mathcal{F} (\tau)$ and $\dot{\mathcal{F}} (\tau)$. 


The matrix elements of $\mathcal{F} (\tau)$ and $\dot{\mathcal{F}} (\tau)$ can be determined using a wide range of numerically exact or approximate propagation methods. 
Since the matrix elements of $\mathcal{F} (\tau)$ and $\dot{\mathcal{F}} (\tau)$ are given in terms of two-time correlation functions of the overall-system,\cite{mulvihill22} the only requirement for a propagation method is that it should be able to calculate such quantities, either exactly or approximately.   

In this paper, we compare and contrast two input methods: the quantum-mechanically exact TT-TFD method described in Sec.~\ref{sec:tt-tfd} and the approximate semiclassical LSCII method, previously described in Ref.~\citenum{mulvihill22}. The inclusion of the LSCII input method is done for the sake of comparison between the memory kernels and inhomogeneous terms as obtained from an approximate input method with those obtained via an exact input method, with the intent of exploring the main sources of inaccuracy when approximate input methods are used. 

For the LSCII method, we calculate ${\cal F}_{jk,lm}(\tau)$ and $\dot{\cal F}_{jk,lm}(\tau)$ directly as described in Ref.~\citenum{mulvihill22}. For the TT-TFD method, we calculate the $N_e^2 \times N_e^2$ elements of the time evolution operator of the electronic reduced density matrix ${\cal U}(\tau)$ introduced by Eq.~(\ref{eq:U}). Then, ${\cal F}_{jk,lm}(\tau)$ and $\dot{\cal F}_{jk,lm}(\tau)$ are obtained from numerical derivatives according to Eq.~(\ref{eq:FFdotU}). For the results given in this paper, the numerical derivatives were calculated using the second-order finite central difference method available in the NumPy Python library. 

Once the PFIs have been obtained with either TT-TFD or LSCII propagation, the memory kernels and inhomogeneous terms of the GQMEs are calculated via an iterative algorithm that solves the corresponding Volterra equation [see Eqs.~\eqref{eq:K_full}, \eqref{eq:Kvolt_pop}, \eqref{eq:Kvolt_DD}, \eqref{eq:Kvolt_AA}, and \eqref{eq:Ivolt_AA}].\cite{mulvihill19a, mulvihill22} The different types of GQMEs [see Eqs.~\eqref{eq:mGQME},\eqref{gqme_pop}, \eqref{gqme_DD}, and \eqref{gqme_AA}] are then solved numerically for the electronic density matrix elements via a Runge-Kutta fourth-order (RK4) algorithm. 


\subsection{Results}

\begin{figure}[!ht]
\centering
\includegraphics[width=0.5\columnwidth]{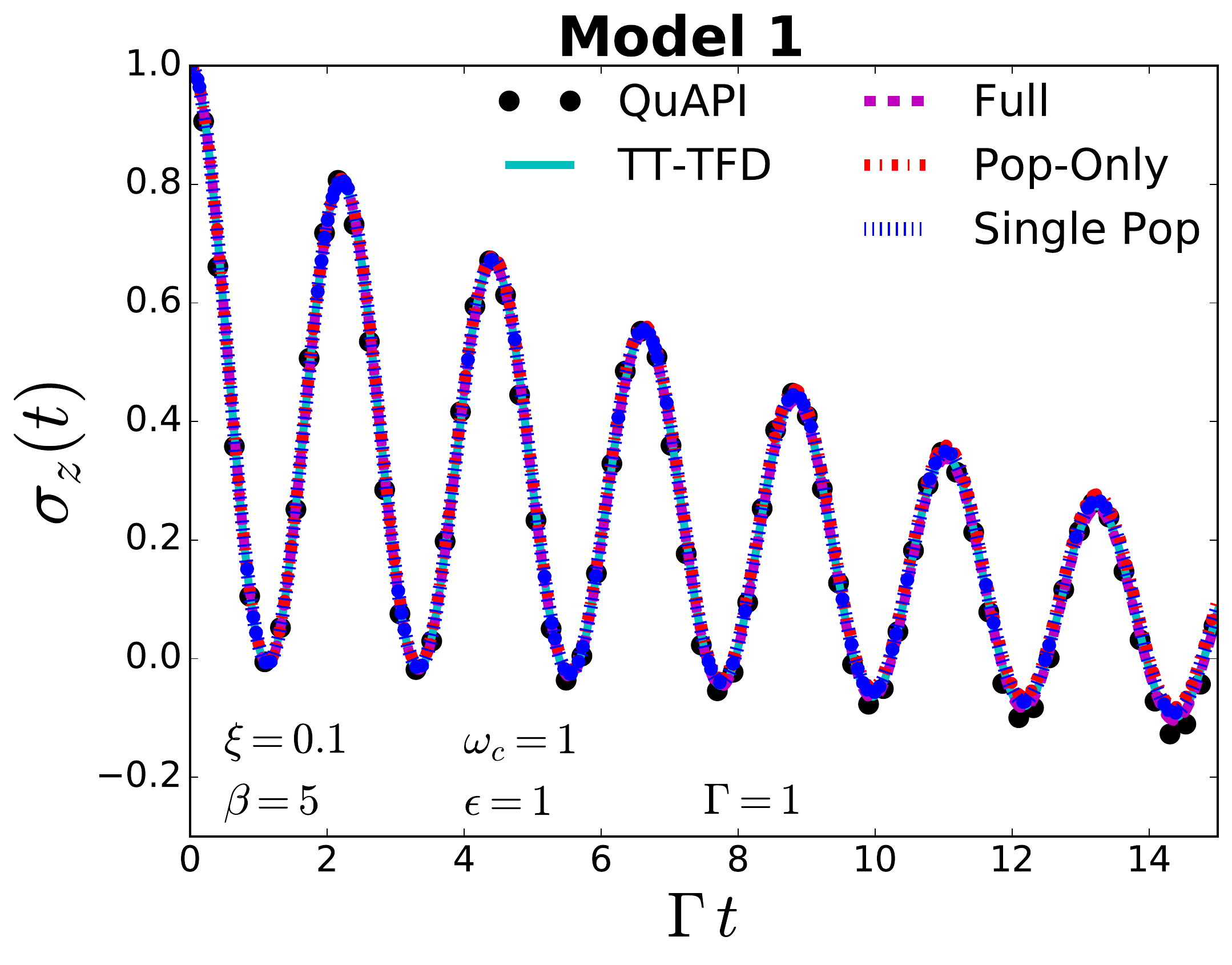} \vspace{-1em}
\caption{Electronic population difference $\sigma_z(t) = \sigma_{DD}(t) - \sigma_{AA}(t)$ as a function of time for model 1 in Table \ref{tab:parameters}.
Shown are exact QuAPI results (black circles) and results obtained based on:
direct application of TT-TFD (solid cyan line); 
a combination of the two single-population scalar GQMEs of the form of Eqs.~\eqref{gqme_DD} and \eqref{gqme_AA} for $\sigma_{DD}(t)$ and $\sigma_{AA}(t)$, respectively, with TT-TFD-based PFIs (dotted blue line);
a populations-only GQME of the form of Eq.~(\ref{gqme_pop}) with TT-TFD-based PFIs (dashed-dotted red line); 
and the full density matrix GQME of the form of Eq.~(\ref{eq:mGQME}) with TT-TFD-based PFIs (dashed magenta line).
}
\label{fig:mod1}
\end{figure}
\begin{figure}[!ht]
\centering
\includegraphics[width=0.5\columnwidth]{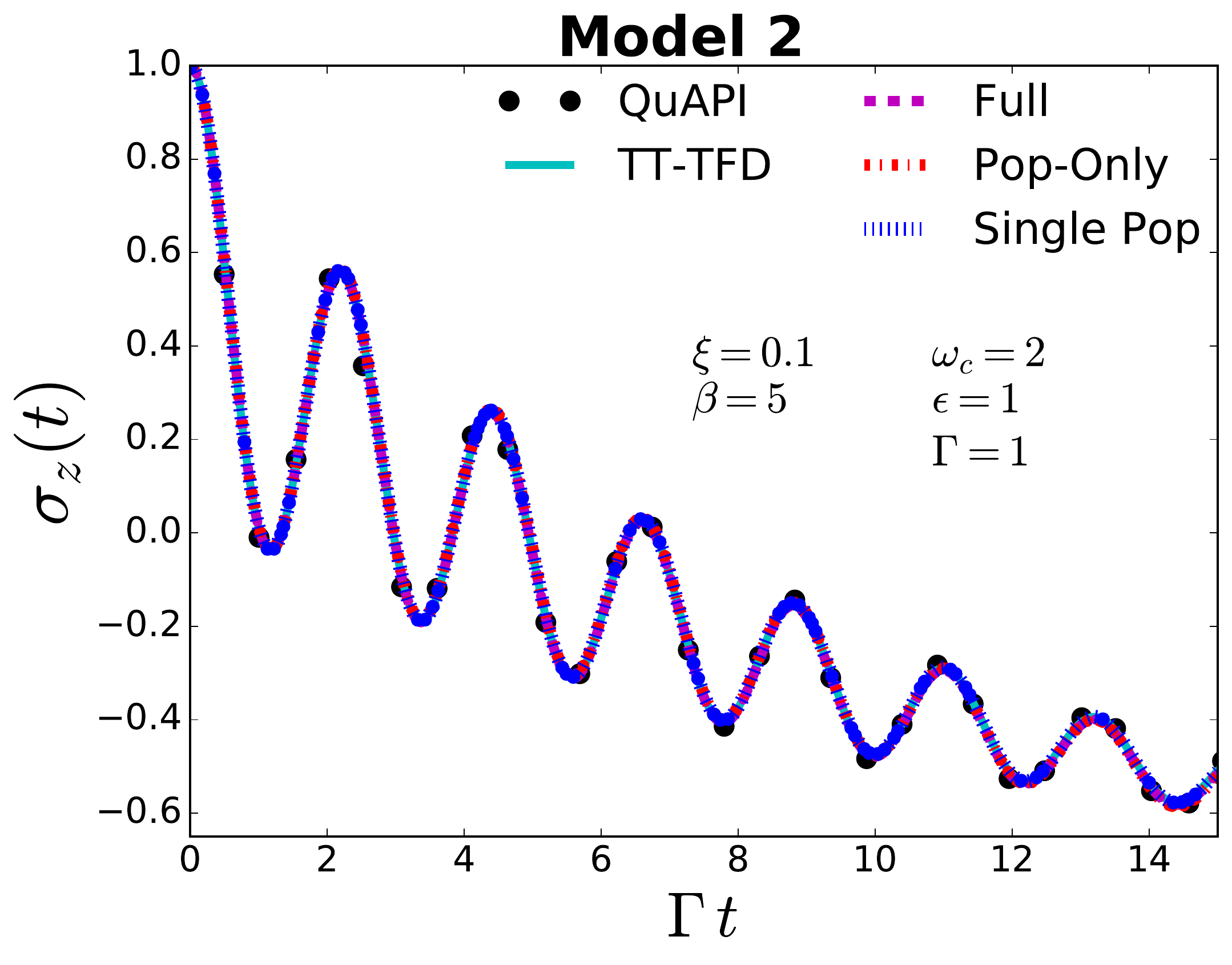} \vspace{-1em}
\caption{Electronic population difference $\sigma_z(t) = \sigma_{DD}(t) - \sigma_{AA}(t)$ as a function of time for model 2 in Table \ref{tab:parameters}.
Shown are exact QuAPI results (black circles) and results obtained based on:
direct application of TT-TFD (solid cyan line); 
a combination of the two single-population scalar GQMEs of the form of Eqs.~\eqref{gqme_DD} and \eqref{gqme_AA} for $\sigma_{DD}(t)$ and $\sigma_{AA}(t)$, respectively, with TT-TFD-based PFIs (dotted blue line);
a populations-only GQME of the form of Eq.~(\ref{gqme_pop}) with TT-TFD-based PFIs (dashed-dotted red line); 
and the full density matrix GQME of the form of Eq.~(\ref{eq:mGQME}) with TT-TFD-based PFIs (dashed magenta line).
}
\label{fig:mod2}
\end{figure}
\begin{figure}[!ht]
\centering
\includegraphics[width=0.5\columnwidth]{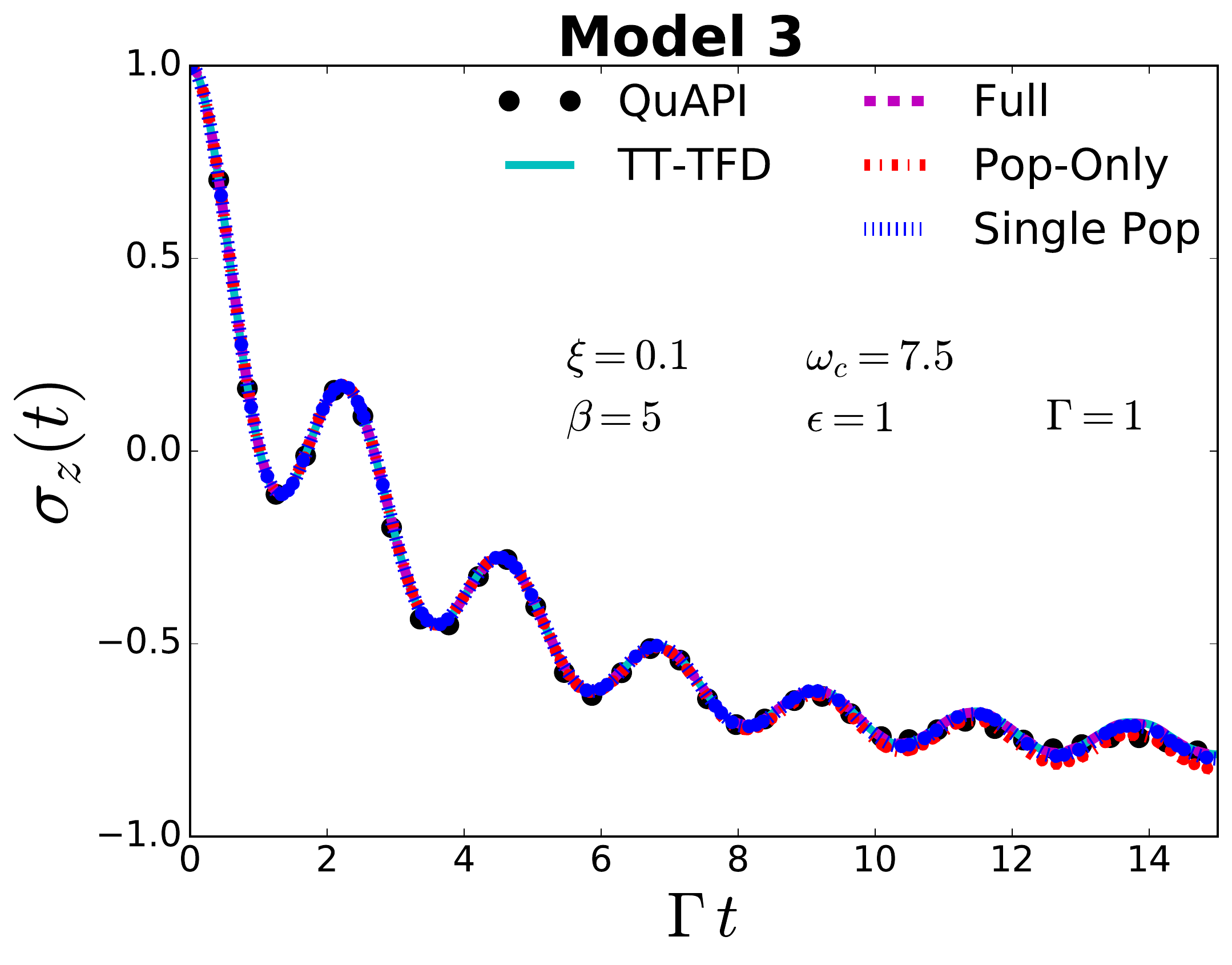} \vspace{-1em}
\caption{Electronic population difference $\sigma_z(t) = \sigma_{DD}(t) - \sigma_{AA}(t)$ as a function of time for model 3 in Table \ref{tab:parameters}.
Shown are exact QuAPI results (black circles) and results obtained based on:
direct application of TT-TFD (solid cyan line); 
a combination of the two single-population scalar GQMEs of the form of Eqs.~\eqref{gqme_DD} and \eqref{gqme_AA} for $\sigma_{DD}(t)$ and $\sigma_{AA}(t)$, respectively, with TT-TFD-based PFIs (dotted blue line);
a populations-only GQME of the form of Eq.~(\ref{gqme_pop}) with TT-TFD-based PFIs (dashed-dotted red line); 
and the full density matrix GQME of the form of Eq.~(\ref{eq:mGQME}) with TT-TFD-based PFIs (dashed magenta line).
}
\label{fig:mod3}
\end{figure}
\begin{figure}[!ht]
\centering
\includegraphics[width=0.5\columnwidth]{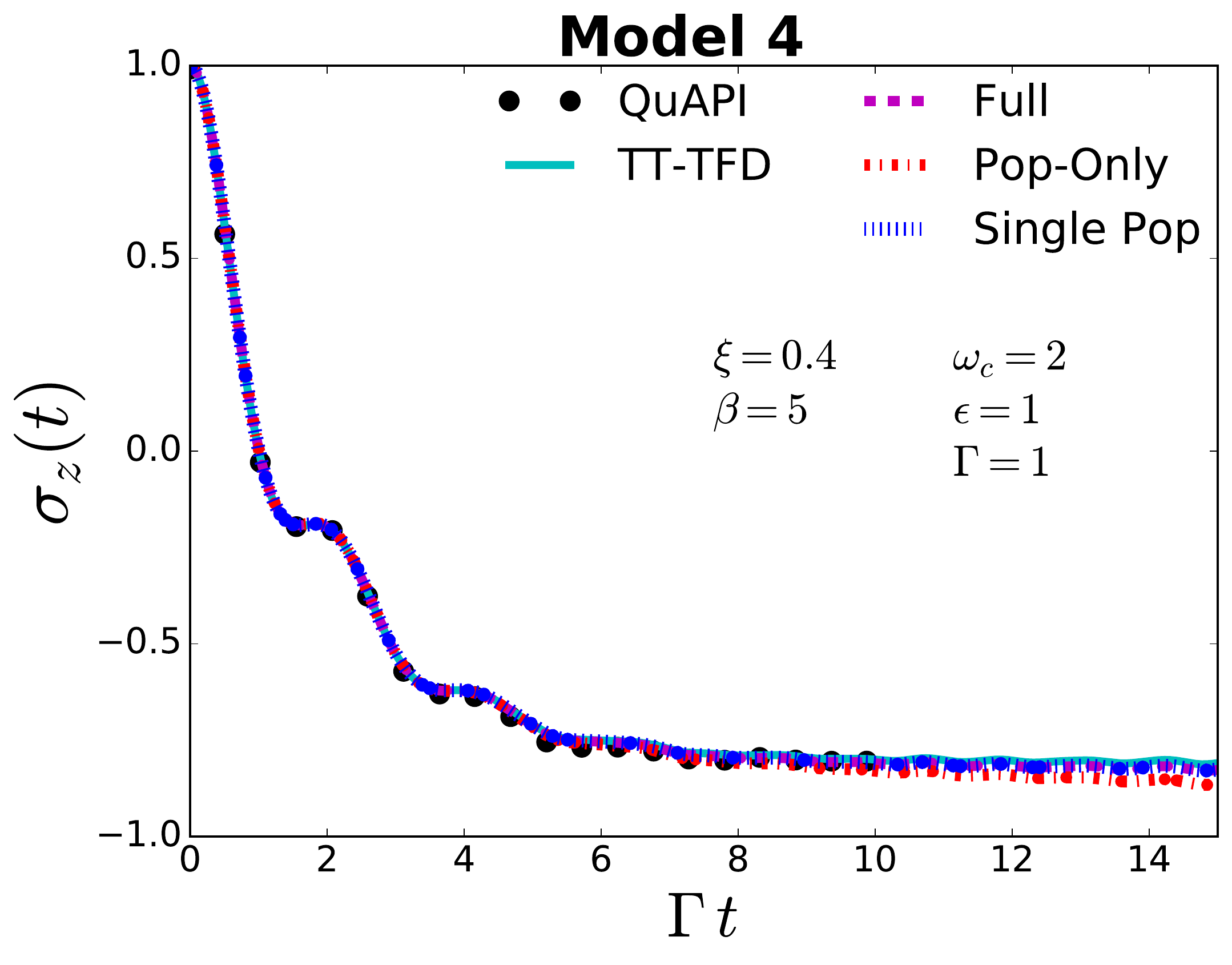} \vspace{-1em}
\caption{Electronic population difference $\sigma_z(t) = \sigma_{DD}(t) - \sigma_{AA}(t)$ as a function of time for model 4 in Table \ref{tab:parameters}.
Shown are exact QuAPI results (black circles) and results obtained based on:
direct application of TT-TFD (solid cyan line); 
a combination of the two single-population scalar GQMEs of the form of Eqs.~\eqref{gqme_DD} and \eqref{gqme_AA} for $\sigma_{DD}(t)$ and $\sigma_{AA}(t)$, respectively, with TT-TFD-based PFIs (dotted blue line);
a populations-only GQME of the form of Eq.~(\ref{gqme_pop}) with TT-TFD-based PFIs (dashed-dotted red line); 
and the full density matrix GQME of the form of Eq.~(\ref{eq:mGQME}) with TT-TFD-based PFIs (dashed magenta line).
}
\label{fig:mod4}
\end{figure}
\begin{figure}[!ht]
\centering
\includegraphics[width=0.5\textwidth]{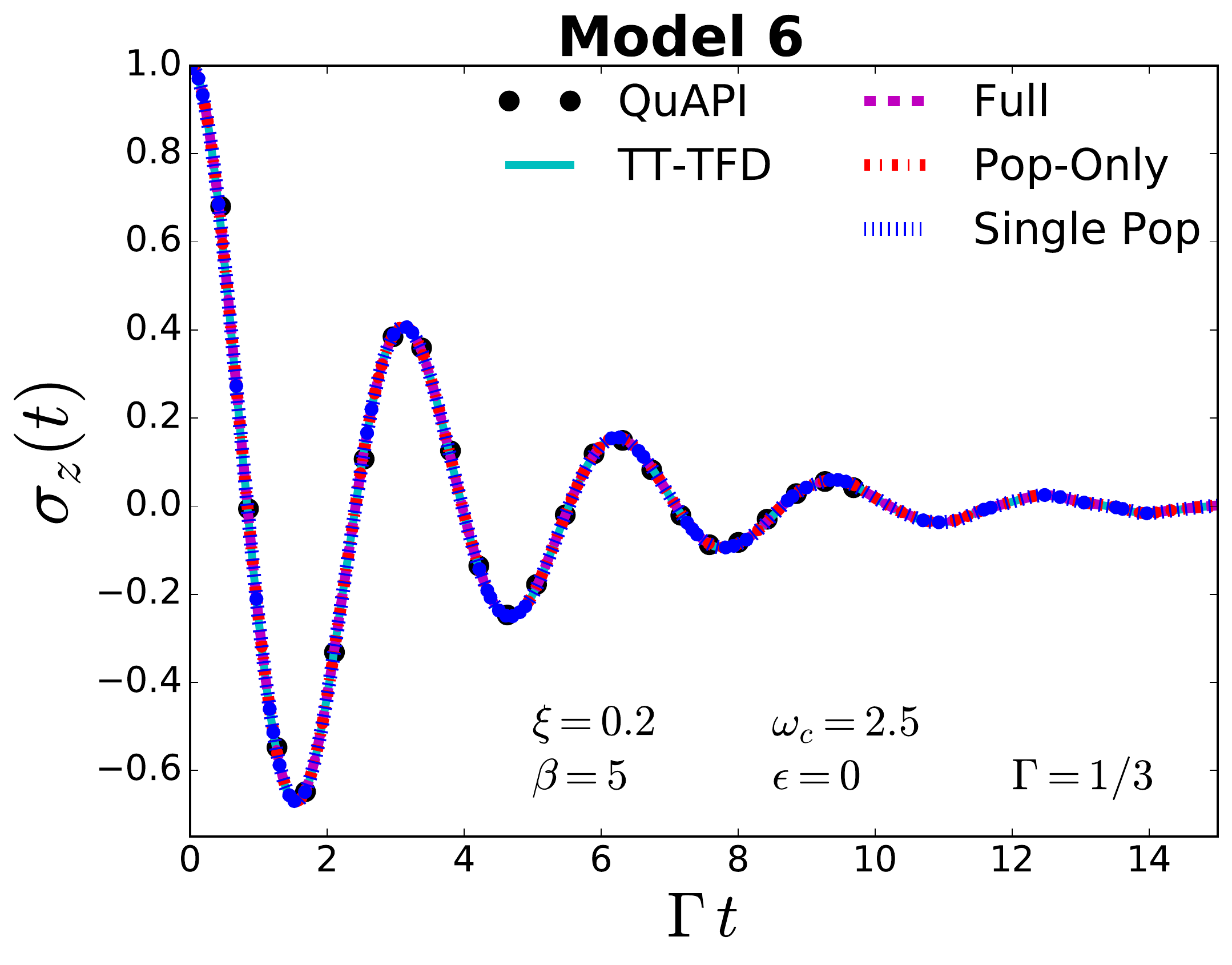} \vspace{-1em}
\caption{Electronic population difference $\sigma_z(t) = \sigma_{DD}(t) - \sigma_{AA}(t)$ as a function of time for model 6 in Table \ref{tab:parameters}.
Shown are exact QuAPI results (black circles) and results obtained based on:
direct application of TT-TFD (solid cyan line); 
a combination of the two single-population scalar GQMEs of the form of Eqs.~\eqref{gqme_DD} and \eqref{gqme_AA} for $\sigma_{DD}(t)$ and $\sigma_{AA}(t)$, respectively, with TT-TFD-based PFIs (dotted blue line);
a populations-only GQME of the form of Eq.~(\ref{gqme_pop}) with TT-TFD-based PFIs (dashed-dotted red line); 
and the full density matrix GQME of the form of Eq.~(\ref{eq:mGQME}) with TT-TFD-based PFIs (dashed magenta line).
}
\label{fig:mod6}
\end{figure}

Figs.\ \ref{fig:mod1}-\ref{fig:mod6} compare the time-dependent  $\sigma_z(t)=\sigma_{DD}(t)-\sigma_{AA}(t)$, showing the differences of electronic populations for the five realizations of the spin-boson model outlined in Sec.~\ref{subsec:spinboson} (see Table \ref{tab:parameters}). These results are obtained by using the four different types of GQMEs outlined in Sec.~\ref{subsec:gqme}, with PFIs computed with the TT-TFD method as described in Sec.~\ref{sec:tt-tfd}. These results provide a clear demonstration of the rather remarkable fact that {\em all four GQMEs correspond to exact equations of motion for the electronic populations} and thereby reproduce the same exact population dynamics when a quantum-mechanically exact input method like TT-TFD is used even though they are quite different in form and dimensionality.


\begin{figure*}[!ht]
\centering
\includegraphics[width=0.4\textwidth]{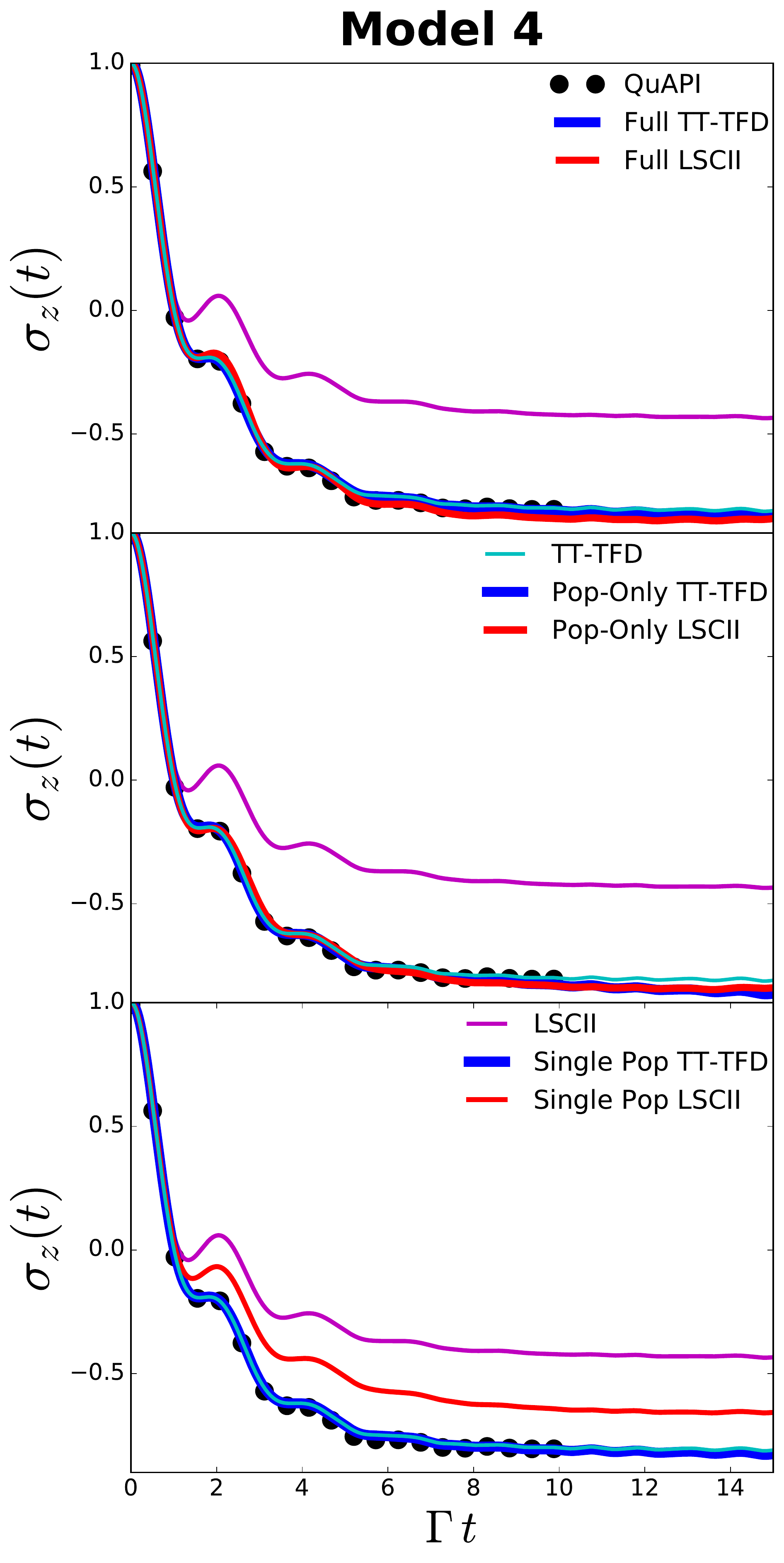} \vspace{-1em}
\caption{Electronic population difference $\sigma_z(t) = \sigma_{DD}(t) - \sigma_{AA}(t)$ as a function of time for model 4 in Table \ref{tab:parameters}. 
Shown are exact QuAPI (black circles) and TT-TFD (green lines) results; LSCII result (purple lines); and full GQME (upper plot), populations-only GQME (middle plot) and combination of two single-population scalar GQMEs (lower plot) results obtained with TT-TFD-based PFIs (blue line) and LSCII based PFIs (red line).
}
\label{fig:sig_Z_mod4}
\end{figure*}

Next, we focus on model 4 for a more detailed analysis, with the analogous analysis for the other models provided in the SI. 
Fig.~\ref{fig:sig_Z_mod4} compares the population relaxation dynamics for model 4 (see Table \ref{tab:parameters}), obtained with different types of GQMEs and memory kernels calculated by TT-TFD and LSCII input methods.
The population relaxation dynamics generated via the LSCII-based populations-only GQME is in excellent agreement with the exact results. At the same time, the population relaxation dynamics generated via the LSCII-based single-population GQMEs is inaccurate. The origin of this discrepancy can be traced back to the fact that the LSCII-based single-population GQMEs do not conserve population (i.e., $\sigma_{DD}(t) + \sigma_{AA}(t) \neq 1$). 

\begin{figure*}[!ht]
\centering
\includegraphics[width=0.95\textwidth]{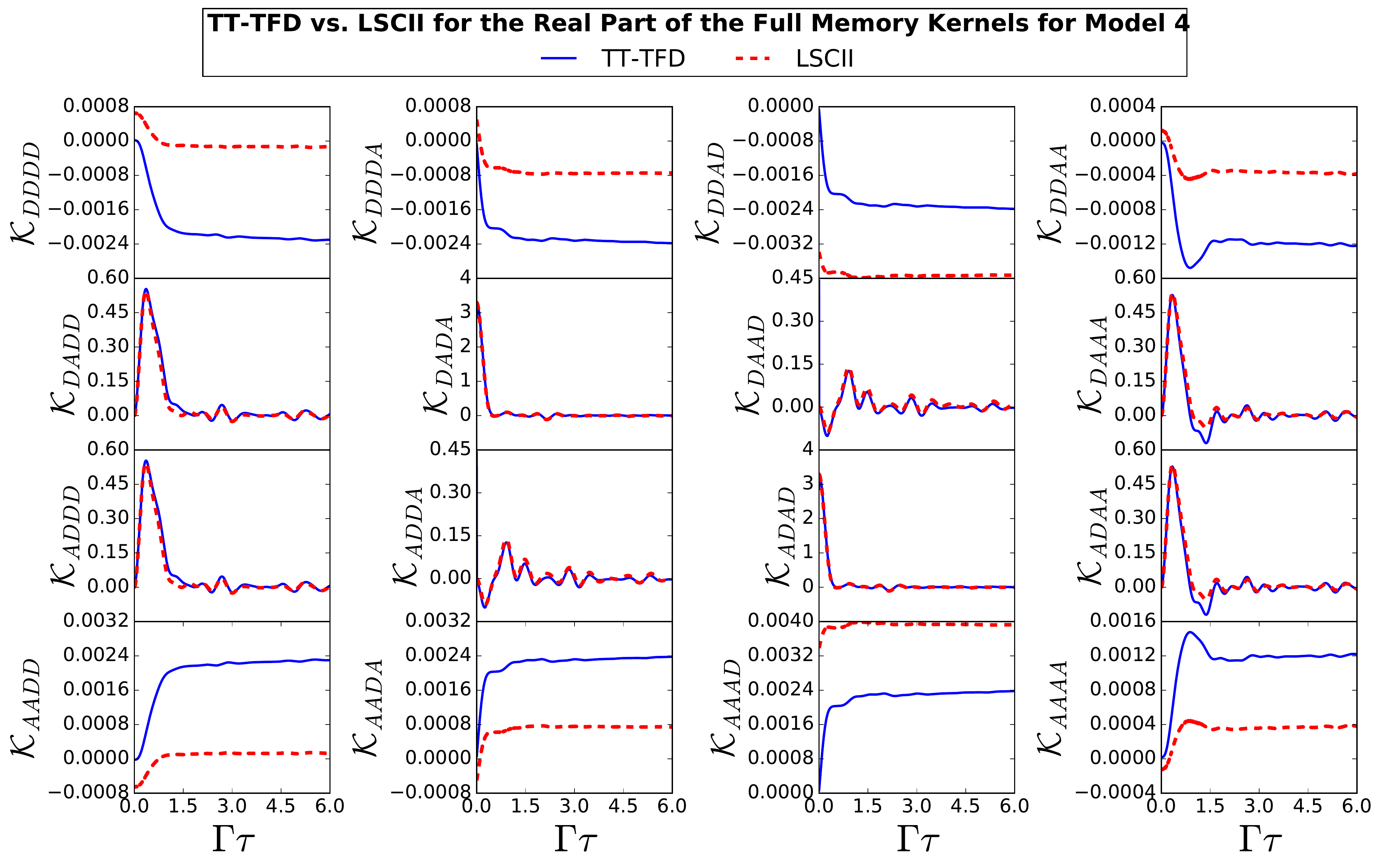} \vspace{-.5em}
\caption{
Real parts of the matrix elements of the memory kernel of the GQME for the full electronic density matrix [${\cal K}^{\text{full}}(\tau)$ in Eq.~(\ref{eq:K_full})] for model 4 as obtained from TT-TFD-based PFIs (solid blue lines) and LSCII-based PFIs (dashed red lines). 
Similar graphs for the other four models are provided in the SI.
}
\label{fig:K4real}
\end{figure*}

\begin{figure*}[!ht]
\centering
\includegraphics[width=0.95\textwidth]{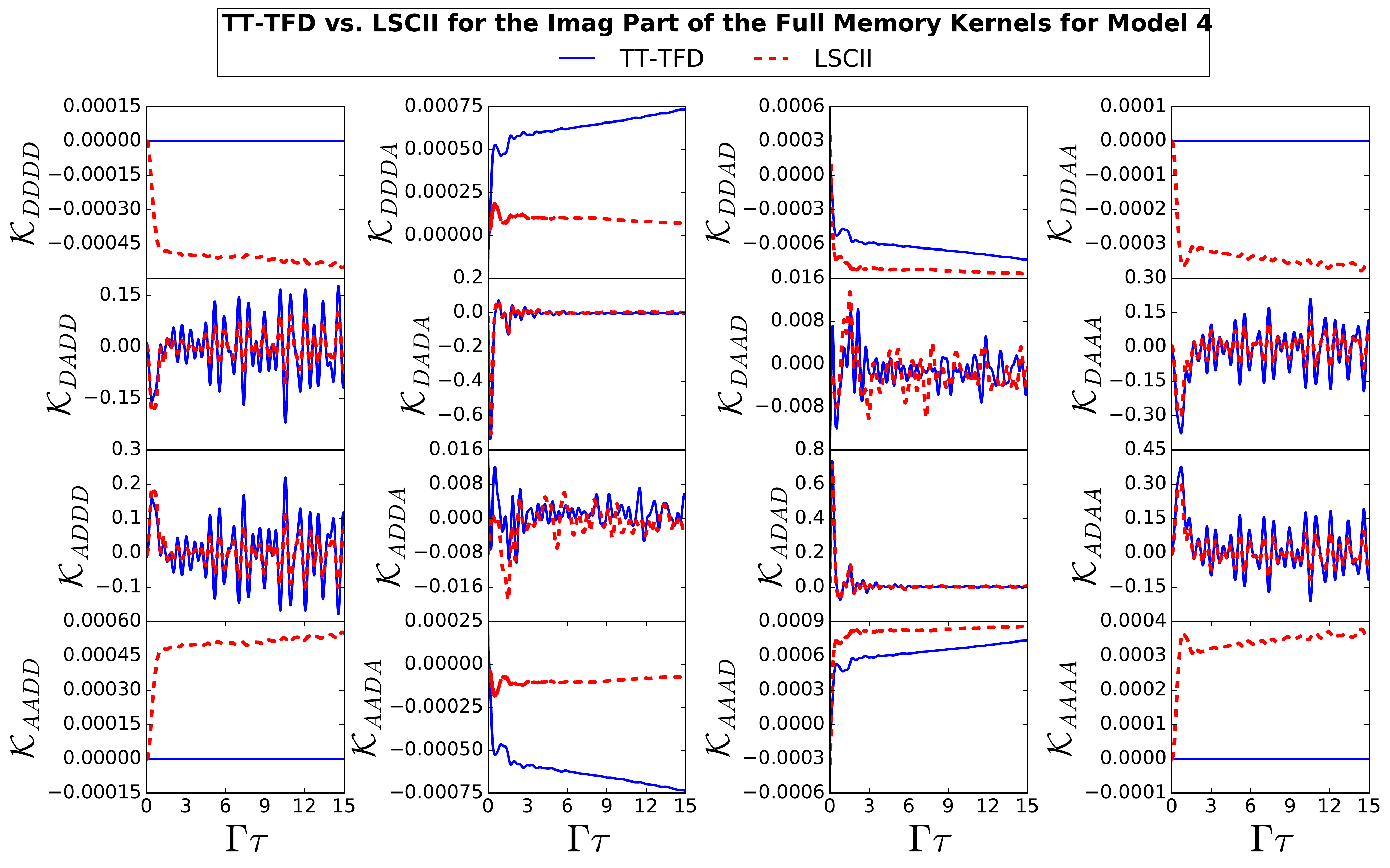} \vspace{-.5em}
\caption{
Imaginary parts of the matrix elements of the memory kernel of the GQME for the full electronic density matrix [${\cal K}^{\text{full}}(\tau)$ in Eq.~(\ref{eq:K_full})] for model 4 as obtained from TT-TFD-based PFIs (solid blue lines) and LSCII-based PFIs (dashed red lines). 
Similar graphs for the other four models are provided in the SI.
}
\label{fig:K4imag}
\end{figure*}

\begin{figure*}[!ht]
\centering
\includegraphics[width=0.9\textwidth]{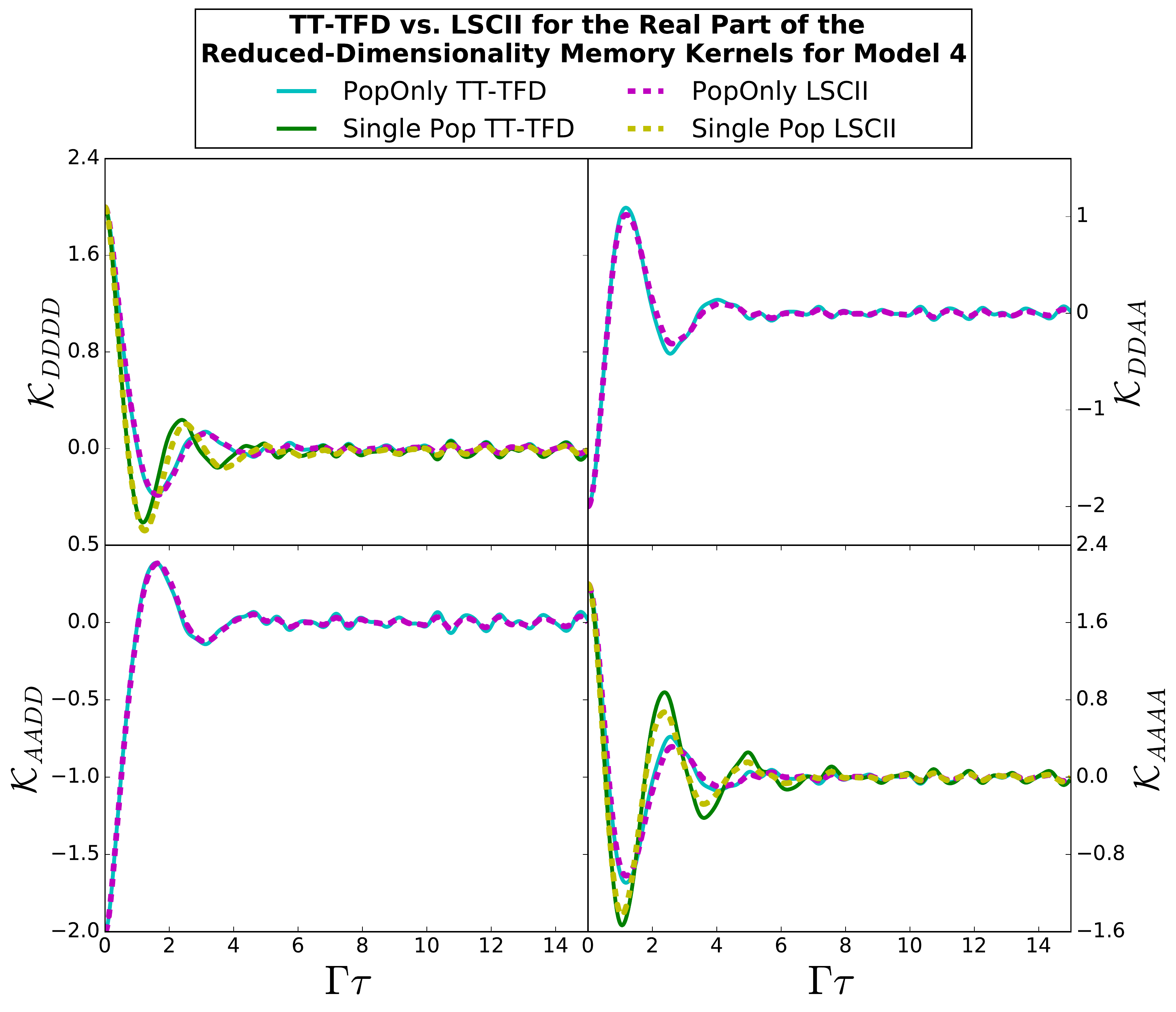} \vspace{-.5em}
\caption{
The real parts of the matrix elements of the memory kernels for the populations-only and single-population GQMEs for model 4 as obtained from TT-TFD-based PFIs and LSCII-based PFIs. 
Shown are the matrix elements of three different memory kernels:
(1) The memory kernel of the populations-only GQME [${\cal K}^{\text{pop}}(\tau)$ in Eq.~(\ref{eq:Kvolt_pop})], 
which has four elements ($ DDDD, DDAA, AADD, AAAA $) and is depicted with solid cyan lines for the results from TT-TFD-based PFIs and dashed magenta lines for the results from LSCII-based PFIs;
(2) and (3) The single-element memory kernels of the scalar single-population GQMEs [${\cal K}^{\text{donor}}_{DD,DD}(\tau)$ and ${\cal K}^{\text{acceptor}}_{AA,AA}(\tau)$, in Eqs.~(\ref{eq:Kvolt_DD}) and \eqref{eq:Kvolt_AA}, respectively], 
which are depicted in the $DDDD$ and $AAAA$ panels, respectively,
with solid green lines for the results from TT-TFD-based PFIs and dashed yellow lines for the results from LSCII-based PFIs.
Graphs with the results for the other four models are provided in the SI.
}
\label{fig:K4real_rd}
\end{figure*}

\begin{figure*}[!ht]
\centering
\includegraphics[width=0.9\textwidth]{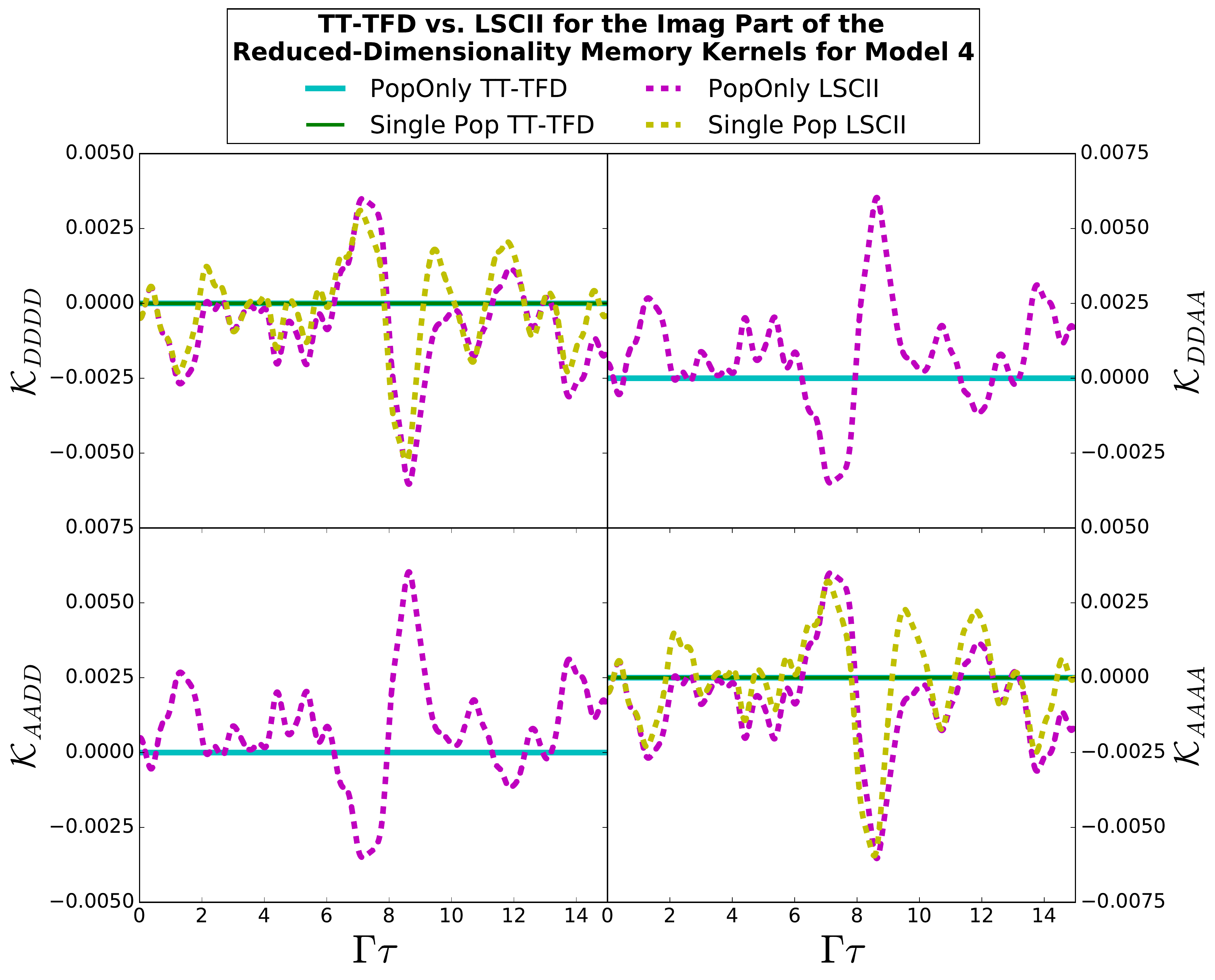} \vspace{-.5em}
\caption{
The imaginary parts of the matrix elements of the memory kernels for the populations-only and single-population GQMEs for model 4 as obtained from TT-TFD-based PFIs and LSCII-based PFIs. 
Shown are the matrix elements of three different memory kernels:
(1) The memory kernel of the populations-only GQME [${\cal K}^{\text{pop}}(\tau)$ in Eq.~(\ref{eq:Kvolt_pop})], 
which has four elements ($ DDDD, DDAA, AADD, AAAA $) and is depicted with solid cyan lines for the results from TT-TFD-based PFIs and dashed magenta lines for the results from LSCII-based PFIs;
(2) and (3) The single-element memory kernels of the scalar single-population GQMEs [${\cal K}^{\text{donor}}_{DD,DD}(\tau)$ and ${\cal K}^{\text{acceptor}}_{AA,AA}(\tau)$, in Eqs.~(\ref{eq:Kvolt_DD}) and \eqref{eq:Kvolt_AA}, respectively], 
which are depicted in the $DDDD$ and $AAAA$ panels, respectively,
with solid green lines for the results from TT-TFD-based PFIs and dashed yellow lines for the results from LSCII-based PFIs.
Graphs with the results for the other four models are provided in the SI.
}
\label{fig:K4imag_rd}
\end{figure*}

Figs.~\ref{fig:K4real} and \ref{fig:K4imag} show the real and imaginary parts of the TT-TFD memory kernels for model 4, as compared to the real and imaginary parts of the LSCII memory kernels for the same model.\cite{mulvihill22}
Each figure includes 16 graphs, corresponding to the elements of the $4 \times 4$ memory kernel matrix.
Since the memory kernel for the full electronic density matrix GQME is represented by a $4 \times 4$ matrix, it has elements in all 16 graphs in Figs.~\ref{fig:K4real} and \ref{fig:K4imag}. 
In contrast, the memory kernel of the populations-only GQME is represented by a $2 \times 2$ matrix [see Eqs.~(\ref{gqme_pop}) and (\ref{eq:Kvolt_pop})]. The real and imaginary parts of the four elements of the populations-only memory kernel are shown in Figs.~\ref{fig:K4real_rd} and \ref{fig:K4imag_rd}. The memory kernels of the two single population scalar GQMEs are scalar [see Eqs.~(\ref{gqme_DD})-(\ref{eq:Kvolt_AA})] and their real and imaginary parts are each therefore shown in one graph (the top left corner for the donor single-population GQME and the bottom right corner for the acceptor single-population GQME in Figs.~\ref{fig:K4real_rd} and \ref{fig:K4imag_rd}, respectively). 

We start the analysis with the memory kernel in the case of the GQME for the full electronic density matrix, ${\cal K}^{\text{full}} (\tau)$.
In this case, all the corner memory kernel elements (${\cal K}^{\text{full}}_{DDDD} $, ${\cal K}^{\text{full}}_{DDAA}$, ${\cal K}^{\text{full}}_{AADD}$ and ${\cal K}^{\text{full}}_{AAAA}$) can be shown to vanish for the TT-TFD-based memory kernels. 
This implies that, in this case, the memory kernel does not give rise to direct coupling terms between populations but population transfer is mediated by coherences. More specifically, within this GQME type, population transfer from the donor ($\sigma_{DD}$) to the acceptor ($\sigma_{AA}$) corresponds to a two-step process. It starts with population-to-coherence transfer induced by coupling between $\sigma_{DD}$ and $\sigma_{DA}$ or $\sigma_{AD}$ and then proceeds to coherence-to-population transfer induced by coupling between $\sigma_{DA}$ or $\sigma_{AD}$ and $\sigma_{AA}$. 

Comparison of the eight matrix elements of the memory kernel that couple populations and coherences, namely $\{ {\cal K}^{\text{full}}_{DADD},{\cal K}^{\text{full}}_{DDDA},{\cal K}^{\text{full}}_{ADDD},{\cal K}^{\text{full}}_{DDAD},{\cal K}^{\text{full}}_{DAAA},{\cal K}^{\text{full}}_{AADA},{\cal K}^{\text{full}}_{ADAA},{\cal K}^{\text{full}}_{AAAD} \}$, reveals several trends:
\begin{itemize}
    \item The agreement between TT-TFD and LSCII is significantly better 
    for the matrix elements $\{ {\cal K}^{\text{full}}_{DADD},{\cal K}^{\text{full}}_{ADDD},{\cal K}^{\text{full}}_{DAAA},{\cal K}^{\text{full}}_{ADAA} \}$ than for the matrix elements 
    $\{ {\cal K}^{\text{full}}_{DDDA},{\cal K}^{\text{full}}_{DDAD},{\cal K}^{\text{full}}_{AADA},{\cal K}^{\text{full}}_{AAAD} \}$.
    At the same time, the four matrix elements $\{ {\cal K}^{\text{full}}_{DDDA},{\cal K}^{\text{full}}_{DDAD},{\cal K}^{\text{full}}_{AADA},{\cal K}^{\text{full}}_{AAAD} \}$ are significantly smaller than the remaining four matrix elements $\{ {\cal K}^{\text{full}}_{DADD},{\cal K}^{\text{full}}_{ADDD},{\cal K}^{\text{full}}_{DAAA},{\cal K}^{\text{full}}_{ADAA} \}$. 
    Thus, LSCII appears to capture the larger-amplitude matrix elements better 
    than the smaller ones. Given the expectation that the larger-amplitude matrix elements would play a more significant role in the dynamics, this observation is consistent with the relative accuracy of the LSCII-based GQME.
    \item Whereas the real parts of the larger matrix elements $\{ {\cal K}^{\text{full}}_{DADD},{\cal K}^{\text{full}}_{ADDD},{\cal K}^{\text{full}}_{DAAA},{\cal K}^{\text{full}}_{ADAA} \}$ are seen to be relatively short-lived (compared to the population relaxation time scale, see Figs.~\ref{fig:mod1}-\ref{fig:mod6}) and exhibit a monotonic decay, the imaginary parts are seen to be oscillatory and do not appear to decay. 
    It should be noted that the oscillatory behavior of the imaginary parts obtained via LSCII is damped compared to exact results obtained via TT-TFD. The observed damping is likely a manifestation of the quasiclassical nature of LSCII, which limits its ability to accurately capture coherent quantum dynamics. Since one expects the real parts to dominate population relaxation rates, 
    the relative accuracy of the LSCII-based GQME can be attributed to the ability of LSCII to capture the real parts rather well. 
\end{itemize}

Examination of the remaining nonvanishing matrix elements, $\{ {\cal K}^{\text{full}}_{DADA},{\cal K}^{\text{full}}_{DAAD},{\cal K}^{\text{full}}_{ADDA},{\cal K}^{\text{full}}_{ADAD} \}$, reveals the following trends:
\begin{itemize}
    \item The real parts of ${\cal K}^{\text{full}}_{DADA}$ and ${\cal K}^{\text{full}}_{ADAD}$ are significantly larger 
    and less oscillatory than the real parts of ${\cal K}^{\text{full}}_{DAAD}$ and ${\cal K}^{\text{full}}_{ADDA}$. This implies that the dynamics of the coherences $\sigma_{DA}$ and $\sigma_{AD}$ is dominated by dephasing (with rates dictated by ${\cal K}^{\text{full}}_{DADA}$ and ${\cal K}^{\text{full}}_{ADAD}$) and that coherence-to-coherence transfer (with rates dictated by ${\cal K}^{\text{full}}_{DAAD}$ and ${\cal K}^{\text{full}}_{ADDA}$) is significantly slower than dephasing. This is consistent with the secular approximation (also called rotating wave approximation), which is often invoked to eliminate coherence transfer terms from perturbative quantum master equations.\cite{baiz11} 
    \item LSCII appears to capture the real parts of $\{ {\cal K}^{\text{full}}_{DADA},{\cal K}^{\text{full}}_{ADAD},{\cal K}^{\text{full}}_{DAAD},{\cal K}^{\text{full}}_{ADDA} \}$ rather accurately. LSCII also appears to be less accurate when it comes to capturing the corresponding imaginary parts, with the inaccuracy manifested by an over-damping of the oscillatory behavior.
    This behavior is similar to that noted above regarding other matrix elements and is consistent with the quasiclassical nature of the approximations on which LSCII is based. 
\end{itemize}
Given that population transfer is mediated by the coherences in the case of the full density matrix GQME, the accuracy of the real parts of the LSCII-based $\{ {\cal K}^{\text{full}}_{DADA},{\cal K}^{\text{full}}_{DAAD},{\cal K}^{\text{full}}_{ADDA},{\cal K}^{\text{full}}_{ADAD} \}$ likely plays an important role in the ability of the LSCII-based GQME to accurately predict the population relaxation dynamics (see Figs.~\ref{fig:mod1} - \ref{fig:mod6}). 


We next consider the memory kernel in the case of the GQME for the electronic populations, ${\cal K}^{\text{pop}} (\tau)$ [see Eqs.~(\ref{gqme_pop}) and (\ref{eq:Kvolt_pop})].
In this case, the memory kernel is given in terms of 
a $2 \times 2$ matrix that consists of only the corner memory kernel elements in Figs.~\ref{fig:K4real} and \ref{fig:K4imag}:
$\{ {\cal K}_{DDDD}^{\text{pop}},{\cal K}_{DDAA}^{\text{pop}}{\cal K}_{AADD}^{\text{pop}}{\cal K}_{AAAA}^{\text{pop}}  \}$. The dimensionality of ${\cal K}^{\text{pop}}(\tau)$ 
should be contrasted with the ${\cal K}^{\text{full}} (\tau)$, for which the same four matrix elements vanish. Since the coherences have been projected out in this case, for this GQME, the memory kernel gives rise to direct coupling between populations, as opposed to population transfer being mediated by the coherences. As a result, donor-to-acceptor population transfer corresponds to a one-step process.   

Comparison of the TT-TFD-based and LSCII-based real and imaginary parts of $\{ {\cal K}_{DDDD}^{\text{pop}},{\cal K}_{DDAA}^{\text{pop}},{\cal K}_{AADD}^{\text{pop}},{\cal K}_{AAAA}^{\text{pop}}  \}$ reveals the following notable trends:
\begin{itemize}
    \item The real parts of those four memory kernel matrix elements are comparable in size and exhibit a damped oscillatory behavior that is longer-lived than the non-vanishing matrix elements of ${\cal K}^{\text{full}} (\tau)$. 
    This behavior is consistent with previous studies\cite{montoyacastillo16,mulvihill22} and can be traced back to the fact that in this case, the memory kernel also needs to account for the impact of the projected-out electronic coherences on the electronic populations.   
    \item LSCII is highly accurate when it comes to reproducing the real parts of the exact TT-TFD-based $\{ {\cal K}_{DDDD}^{\text{pop}},{\cal K}_{DDAA}^{\text{pop}},{\cal K}_{AADD}^{\text{pop}},{\cal K}_{AAAA}^{\text{pop}}  \}$. Given that the real parts of the ${\cal K}^{\text{pop}} (\tau)$ matrix elements dominate the population transfer kinetics, this observation is consistent with the previously made observation that the LSCII-based populations-only GQME can reproduce the population relaxation rather well.\cite{mulvihill22}
    \item Whereas the imaginary parts of 
    $\{ {\cal K}_{DDDD}^{\text{pop}},{\cal K}_{DDAA}^{\text{pop}},{\cal K}_{AADD}^{\text{pop}},{\cal K}_{AAAA}^{\text{pop}}  \}$ computed with TT-TFD vanish, the corresponding LSCII values do not. 
    The discrepancy is due to errors in the calculation of $\dot{\cal F}_{jj,mm}(\tau)$ elements with LSCII, which generates a small real part for the $\dot{\cal F}_{jj,mm}(\tau)$ elements, which should be purely imaginary.
    However, the failure of LSCII to accurately predict the imaginary parts does not appear to impact the accuracy of the population transfer kinetics since the imaginary parts are two orders of magnitude smaller than the real parts. 
\end{itemize}

Finally, we consider the scalar memory kernels in the donor and acceptor single-population GQMEs, ${\cal K}_{DDDD}^{\text{donor}} (\tau)$ and ${\cal K}_{AAAA}^{\text{acceptor}} (\tau)$, respectively  [see Eqs.~(\ref{gqme_DD})-(\ref{eq:Kvolt_AA})]. In this case, 
${\cal K}_{DDDD}^{\text{donor}} (\tau)$ is given by the top-left corner element and ${\cal K}_{AAAA}^{\text{acceptor}} (\tau)$  is given by the bottom-left corner element in Figs.~\ref{fig:K4real_rd} and \ref{fig:K4imag_rd}. Comparison of the real and imaginary parts of ${\cal K}_{DDDD}^{\text{donor}} (\tau)$ and ${\cal K}_{AAAA}^{\text{acceptor}} (\tau)$ computed with TT-TFD and LSCII reveals the following notable trends:
\begin{itemize}
    \item The real parts of ${\cal K}_{DDDD}^{\text{donor}} (\tau)$ and ${\cal K}_{AAAA}^{\text{acceptor}} (\tau)$ are comparable in size and exhibit a damped oscillatory behavior with a lifetime similar to that of the populations-only memory kernel elements.
    \item LSCII is highly accurate for reproducing the real part of the exact TT-TFD-based ${\cal K}_{DDDD}^{\text{donor}} (\tau)$. The accuracy 
    is somewhat lower for reproducing the real part of ${\cal K}_{AAAA}^{\text{acceptor}} (\tau)$.  
    \item While the imaginary parts of 
    ${\cal K}_{DDDD}^{\text{donor}} (\tau)$ and ${\cal K}_{AAAA}^{\text{acceptor}} (\tau)$ computed with TT-TFD
    vanish, the corresponding LSCII values do not. However, the failure of LSCII to accurately predict the imaginary parts does not appear to impact the accuracy of the population transfer kinetics since the imaginary parts are two orders of magnitude smaller than the real parts. 
\end{itemize}

\begin{figure*}[!ht]
\centering
\includegraphics[width=0.9\textwidth]{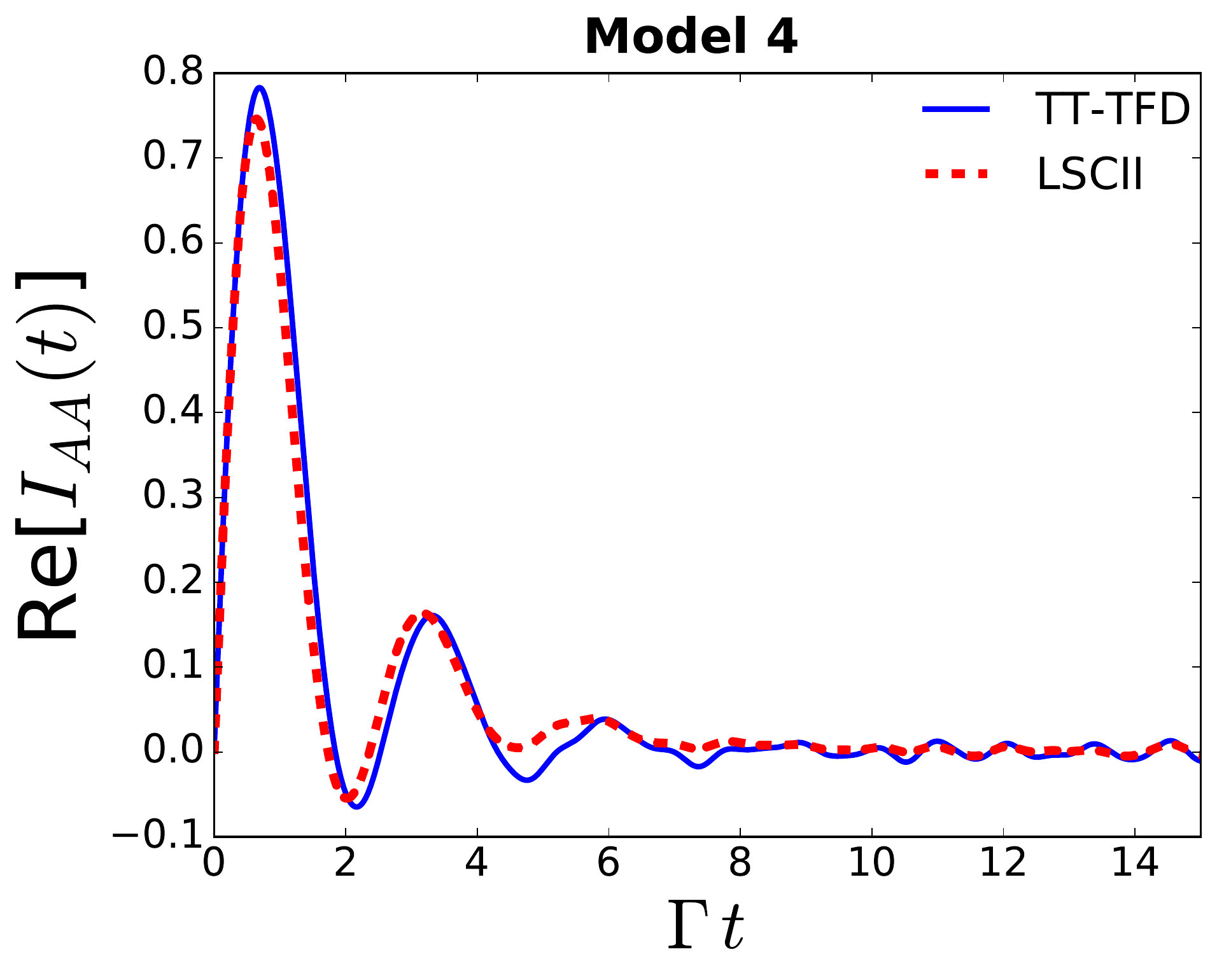} \vspace{-1em}
\caption{
Real part of $\hat{I}_{AA}(\tau)$ [see Eq.~(\ref{eq:Ivolt_AA})] 
for model 4, 
as obtained from TT-TFD-based PFIs (solid blue lines) and LSCII-based PFIs (dashed red lines).
Similar graphs for the other four models are included in the SI.
}
\label{fig:I_mod4}
\end{figure*}

In Fig.~\ref{fig:I_mod4}, we show the real part of the inhomogeneous term of the acceptor single-population GQME, $\hat{I}_{AA}(t)$, which is the only GQME with an inhomogeneous term considered in this paper. The imaginary component is not shown because it is zero for the results from both TT-TFD- and LSCII-based PFIs. In the figure, we see that the inhomogeneous term from LSCII-based PFIs is slightly overdamped compared to the inhomogeneous term from the TT-TFD-based PFIs.

\begin{figure*}[!ht]
\centering
\includegraphics[width=0.9\textwidth]{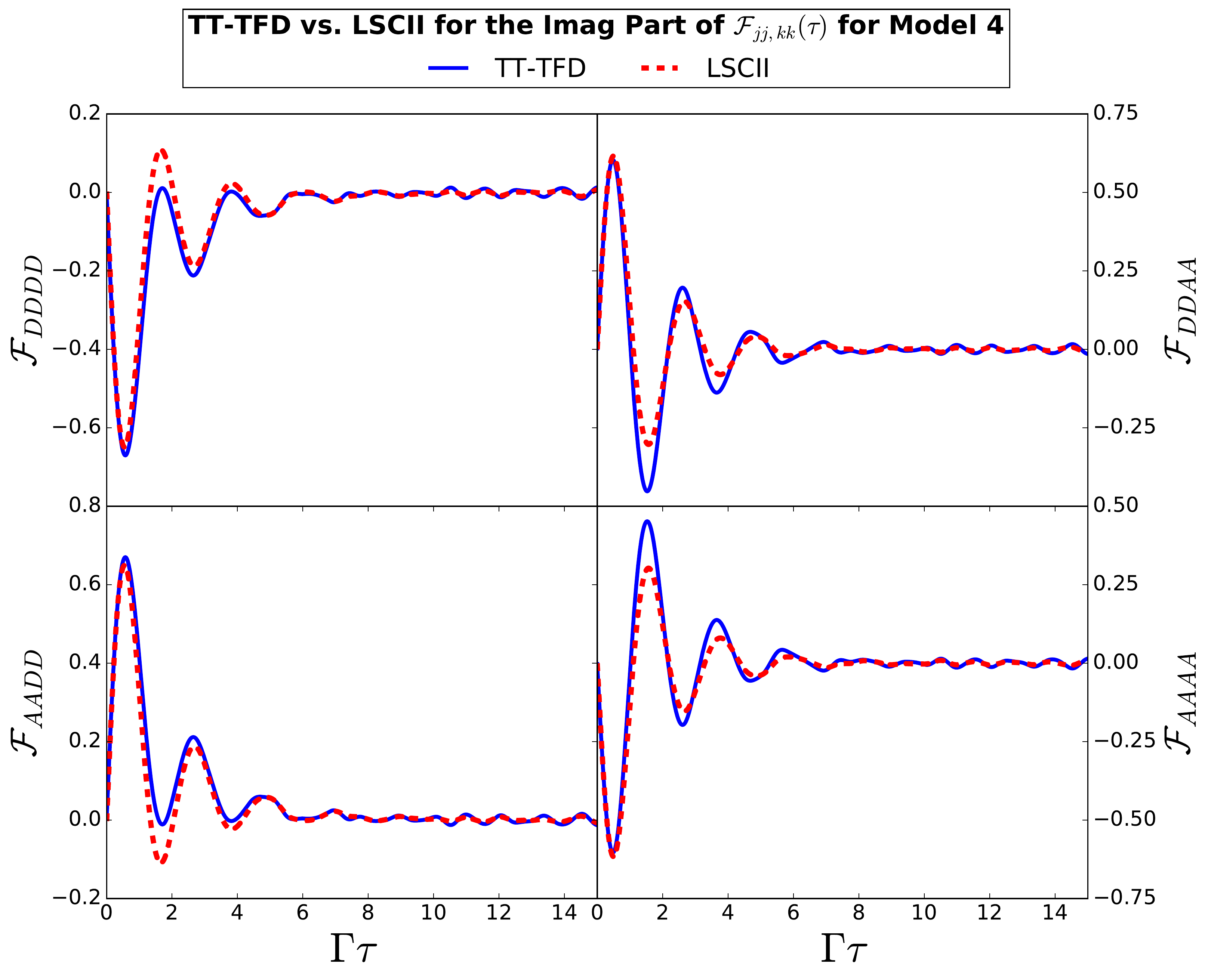} \vspace{-1em}
\caption{
Imaginary parts of the $DDDD$, $DDAA$, $AADD$ and $AAAA$ matrix elements of ${\cal F}(\tau)$ [see Eq.~(\ref{eq:FFdotsub})] 
for model 4, 
as obtained via TT-TFD (solid blue lines) and LSCII (dashed red lines).
Similar graphs for the other four models are included in the SI.
}
\label{fig:Imag_F_mod4}
\end{figure*}

\begin{figure*}[!ht]
\centering
\includegraphics[width=0.9\textwidth]{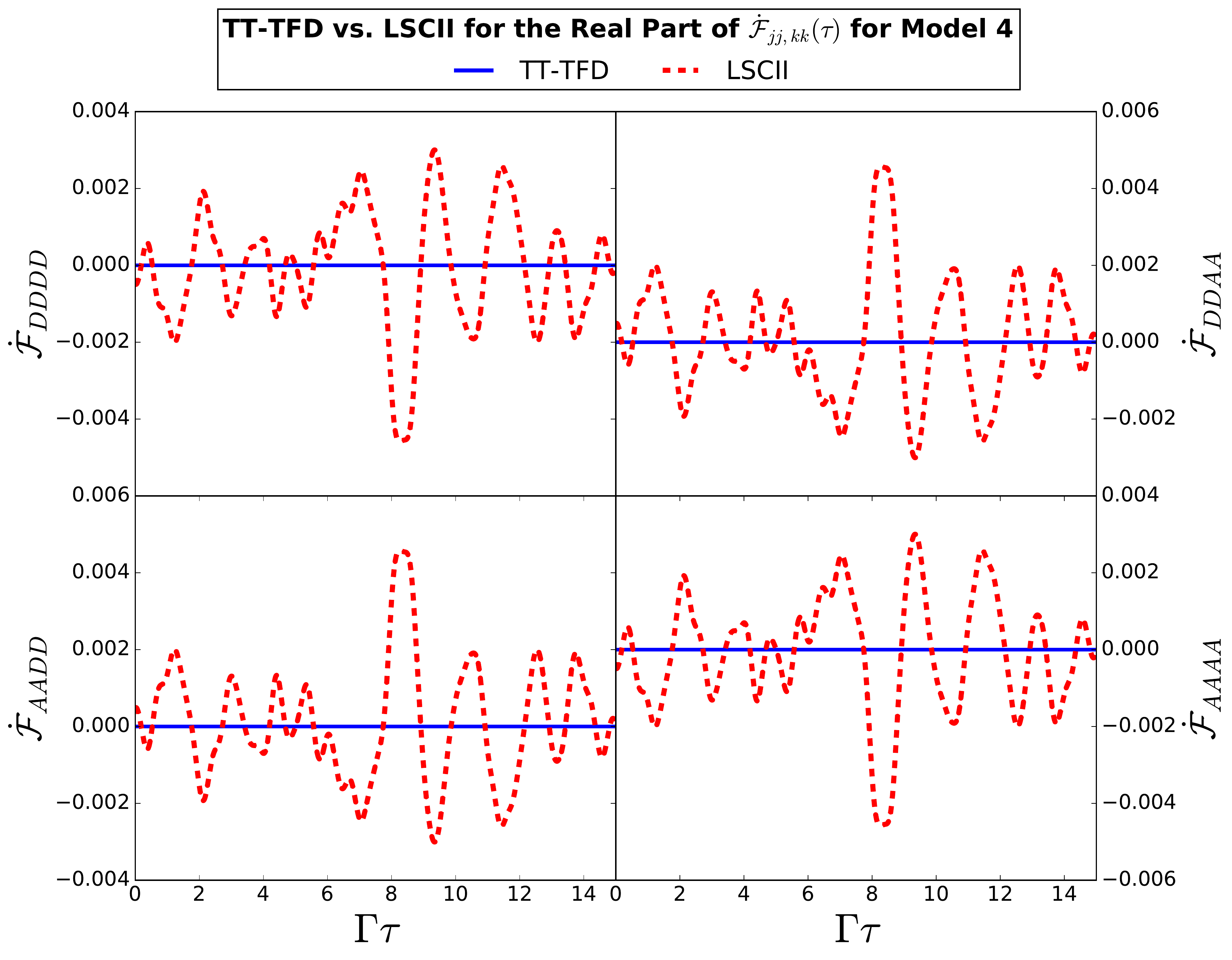}
\caption{
Real parts of the $DDDD$, $DDAA$, $AADD$ and $AAAA$ matrix elements of $\dot{\cal F}(\tau)$ [see Eq.~(\ref{eq:FFdotsub})] 
for model 4, 
as obtained via TT-TFD (solid blue lines) and LSCII (dashed red lines).
Similar graphs for the other four models are included in the SI.
}
\label{fig:Real_Fdot_mod4}
\end{figure*}

\begin{figure*}[!ht]
\centering
\includegraphics[width=0.9\textwidth]{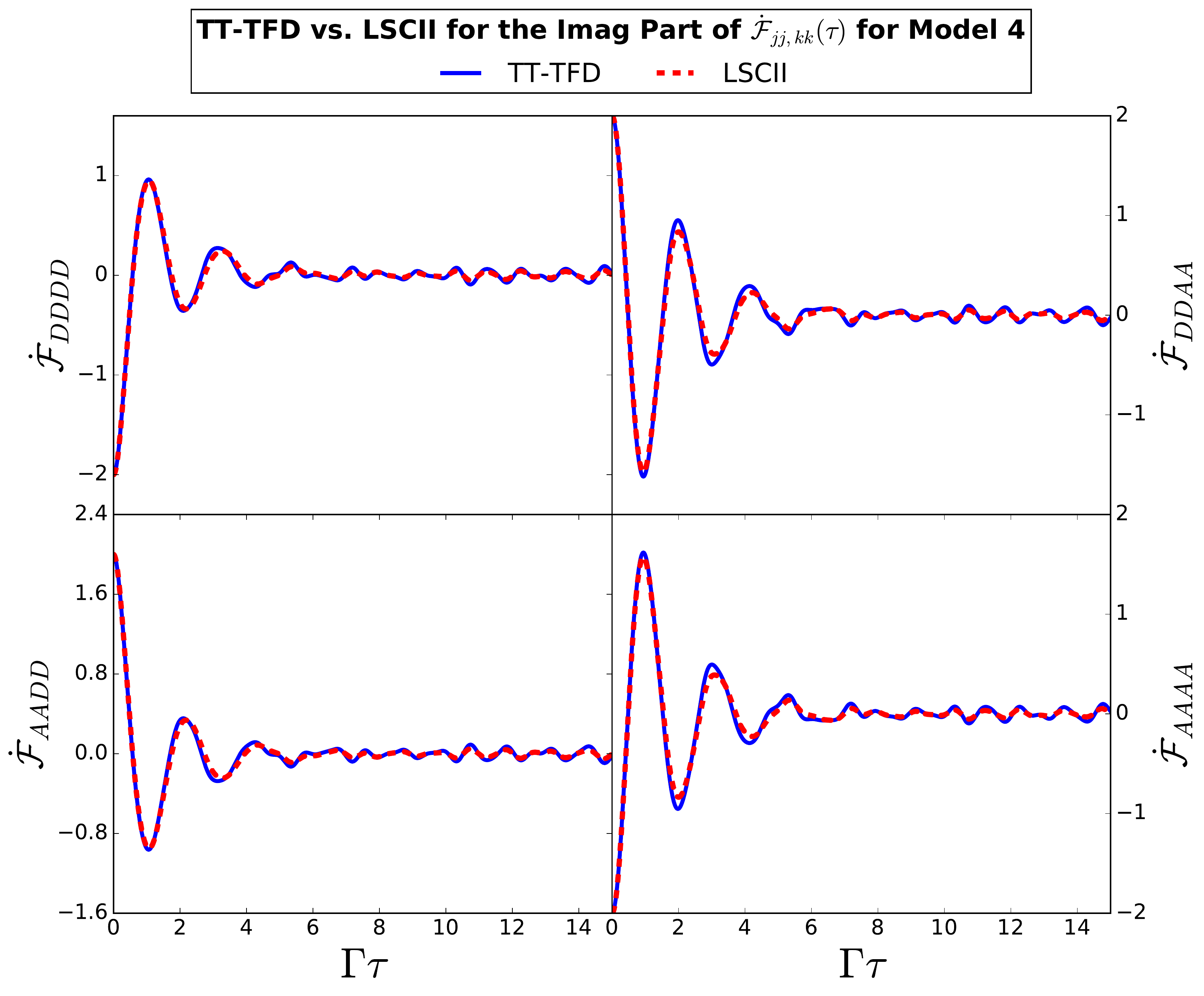} \vspace{-1em}
\caption{
Imaginary parts of the $DDDD$, $DDAA$, $AADD$ and $AAAA$ matrix elements of $\dot{\cal F}(\tau)$ [see Eq.~(\ref{eq:FFdotsub})] 
for model 4, 
as obtained via TT-TFD (solid blue lines) and LSCII (dashed red lines).
Similar graphs for the other four models are included in the SI.
}
\label{fig:Imag_Fdot_mod4}
\end{figure*}

To understand the origin of the inaccuracies in the LSCII-based single-population GQMEs relative to the populations-only GQME, we note that any such inaccuracies must come from inaccuracies in ${\cal F}(\tau)$ and $\dot{\cal F}(\tau)$, as the subsequent steps of the GQME approach are exact. 
To this end, we show in Fig.~\ref{fig:Imag_F_mod4} the imaginary components of the matrix elements of ${\cal F}(\tau)$ and in Figs.~\ref{fig:Real_Fdot_mod4} and \ref{fig:Imag_Fdot_mod4}, the real and imaginary components of the matrix elements of $\dot{\cal F}(\tau)$ that are used as PFIs to obtain the memory kernels for the single-population and populations-only GQMEs.  The real parts of ${\cal F}(\tau)$ are not shown because they are zero for these elements from both LSCII and TT-TFD. These figures clearly show that, although the LSCII-based ${\cal F}(\tau)$ and $\dot{\cal F}(\tau)$ matrix elements can be rather accurate, there are significant deviations from the exact ones. The deviations are the origin of any inaccuracies in the memory kernels obtained from them. 


We now show that, although the errors in ${\cal F} (\tau)$ affect the memory kernels of both single-population and populations-only GQMEs, the effect is weaker on the latter due to error cancellation. 
To see this difference in effect, 
we note that $K_{DD,DD}^{\text{pop}}(\tau)$ and $K_{DD,DD}^{\text{donor}}(\tau)$ are obtained from the PFIs via Eq.~\eqref{eq:Kvolt_sub}:
\begin{align}
K_{DD,DD}^{\text{pop}}(\tau) &= i\dot{\cal F}_{DD,DD}(\tau)+i\int_0^\tau d\tau' \Big[{\cal F}_{DD,DD}(\tau-\tau')K_{DD,DD}^{\text{pop}}(\tau') \label{eq:F0000_poponly}
\\ &\twqq\twqq\twqq+ {\cal F}_{DD,AA}(\tau-\tau')K_{AA,DD}^{\text{pop}}(\tau')\Big], \nonumber
\\ K_{DD,DD}^{\text{donor}}(\tau) &= i\dot{\cal F}_{DD,DD}(\tau)+i\int_0^\tau d\tau' {\cal F}_{DD,DD}(\tau-\tau')K_{DD,DD}^{\text{donor}}(\tau'), \label{eq:F0000_scalar}
\end{align}
where the term that involves the reduced system Liouvillian $\langle \mathcal{L}\rangle_n^0$ is dropped because $\langle \mathcal{L}_{jj,kk}\rangle_n^0 = 0$. 
Importantly, the integrand on the right-hand side of Eq.~\eqref{eq:F0000_poponly} gives rise to inherent error cancellation, since $K_{DD,DD}^{\text{pop}}(\tau)$ and $K_{AA,DD}^{\text{pop}}(\tau)$ are of opposite sign, which causes errors in ${\cal F}(\tau)$ to cancel. On the other hand, Eq.~\eqref{eq:F0000_scalar} does not allow for such error cancellation, thereby making the single-population GQMEs less accurate than the populations-only GQME. 

\subsection{Computational Cost}

In this section, we examine the scaling of the computational cost of the GQME approach with TT-TFD as the input method with respect to GQME type. 

We begin by considering the time step used to calculate the TT-TFD-based PFIs to obtain converged memory kernel and the inhomogeneous term. 
In contrast to LSCII which required a similar time step for all GQMEs,\cite{mulvihill22}, the time step needed for convergence is found to decrease with decreasing dimensionality. 
More specifically, whereas the results shown above are all for a time step of $\Delta t = 0.00150083\, \Gamma^{-1}$, the time step needed for convergence for the full density matrix GQME is in the range of $\Delta t = (0.00300166 - 0.00450249)  \,\Gamma^{-1}$ in contrast to the time step of
$\Delta t = 0.00150083\, \Gamma^{-1}$ required in the case of the populations-only and single-population GQMEs.

In Ref.~\citenum{mulvihill22}, we noted that the direct calculation of $\dot{{\cal F}}_{jk,lm}(\tau)$ given in Eq.~\eqref{eq:FFdotsub} requires calculating the dynamics for more electronic initial conditions than only $|j\rangle \langle k|$ due to terms involving off-diagonal components of the Hamiltonian in the initial state. However, although direct calculation of $\dot{{\cal F}}_{jk,lm}(\tau)$ is necessary when using approximate input methods; when using exact input methods, we can obtain $\dot{{\cal F}}_{jk,lm}(\tau)$ from ${\cal U}_{jk,lm}(\tau)$ as described in Eq.~\eqref{eq:FFdotU}. We therefore only need to calculate the dynamics for the initial electronic state $|j\rangle\langle k|$ to obtain ${\cal U}_{jk,lm}(\tau)$ and subsequently ${\cal F}_{jk,lm}(\tau)$ and $\dot{{\cal F}}_{jk,lm}(\tau)$ through Eq.~\eqref{eq:FFdotU}. 
As a result, there is a significant reduction in the number of initial electronic states necessary for calculating the PFIs needed for the reduced-dimensionality GQMEs compared to the full GQME. More specifically, although the full GQME approach requires simulating the dynamics for 4 initial electronic states in the case of a two-state system, the populations-only GQME requires only 2 initial electronic states, the acceptor single-population GQME approach requires 2 initial electronic states (with one of them due to the inhomogeneous term), and the donor single-population GQME requires only one initial electronic state. Thus, reduced-dimensionality GQMEs significantly enhance computational efficiency with regards to the number of initial states that need to be simulated when exact input methods like TT-TFD are used. 

Next, we consider the cost of obtaining the memory kernels from the PFIs.
The computational complexity of each iteration in the Volterra algorithm for the memory kernel is expected to be $O(N^3_{\text{mat}})$, where $N_{\text{mat}}$ is the number of matrix elements in a row of the memory kernel matrix (e.g., $N_{\text{mat}} = N_e^2$ for the full GQME, $N_{\text{mat}} = N_e$ for the populations-only GQME, and $N_{\text{mat}} = 1$ for the single-population GQMEs). This is true regardless of the input method used and therefore the cost of each iteration of the Volterra algorithm increases dramatically with memory kernel size. The computational complexity of each iteration in the Volterra algorithm for the inhomogeneous term scales more favorably at $O(N^2_{\text{mat}})$ but may still become restrictive with increasing dimensionality. However, it should be noted that the inhomogeneous term often is not needed for the larger-dimensional full and populations-only GQME approaches.
    
The number of iterations required for the iterative Volterra algorithm for the memory kernel to converge is also rather sensitive to the type of GQME and the dimensionality of the electronic observable of interest. More specifically, whereas 2 iterations are required for calculating the single-population memory kernels and 2-3 iterations are needed in the case of the populations-only memory kernel for all the models, 5-7 iterations are required for the full GQME approach. 


An inhomogeneous term is only required for the acceptor single-population GQME approach and would be required for any GQME approach where the set of electronic states that it projects onto does not include the initial electronic state. Because of the scaling of the Volterra algorithm for the inhomogeneous term, it is generally only favorable to use a GQME approach that requires an inhomogeneous term where the dimensionality of the set of electronic states projected onto is small. 

    
The converged memory time for each of the models and GQME types is found using the algorithm outlined in the SI of Ref.~\citenum{mulvihill22}. The basic premise of the algorithm is to first calculate the dynamics at the highest possible memory time, $t_{\text{mem, max}}$, based on the maximum time of the PFI dynamics and then proceed backwards in memory time to find the shortest memory time that keeps each element and time step of the electronic density matrix within a convergence parameter 
when compared to the same element and time step of the dynamics with the highest possible memory time. For the models studied in this paper, the highest possible memory time was $t_{\text{mem, max}} = 15\, \Gamma^{-1}$. The converged memory time for each model and GQME approach is given Table \ref{tab:memtime}. In agreement with the results for LSCII in Ref.~\citenum{mulvihill22}, the full GQME typically corresponds to the shortest memory time and the reduced-dimensionality GQMEs requires significantly longer memory times, particularly the scalar single-population GQMEs. Whereas the RK4 algorithm is expected to have computational complexity $O(t_{\text{mem}})$, the cost of a single iteration of the Volterra algorithm for the memory kernel has quadratic computational complexity $O(t_{\text{mem}}^2)$. Thus, situations where the reduced dimensionality of the electronic observable of interest leads to longer memory time increases computational cost.  

\begin{table}[t]
\centering
\caption{{\bf Memory Time of Each GQME Approach for Each Model}\\ In this table, the colors are  provide a visual aid, with red indicating a memory time above $12\,\Gamma^{-1}$, yellow indicating a memory time from $9 - 12\, \Gamma^{-1}$, and green indicating a memory tie below $9 \Gamma^{-1}$.}
\def\arraystretch{1.125}
\resizebox{1.\columnwidth}{!}{
\begin{tabular}{|c||c||c|c|c|c|}
  \hline
 Model $\#$ & \makecell{Input\\Method} & Full GQME & \makecell{Populations-Only\\GQME} & \makecell{Donor GQME} & \makecell{Acceptor GQME}
  \\ \hline
  \multirow{2}{*}{1} & TT-TFD & \cellcolor{green!22} 5.5034 & \cellcolor{red!20} 14.7534 & \cellcolor{red!20} 14.7534 & \cellcolor{red!20} 14.5034
  \\ \cline{2-6}
     & LSCII & \cellcolor{green!22} 5.25415 & \cellcolor{red!20} 14.7541 & \cellcolor{red!20} 14.5041 & \cellcolor{red!20} 14.5041
  \\ \hline\hline
  \multirow{2}{*}{2} & TT-TFD & \cellcolor{green!22} 1.65348 & \cellcolor{red!20} 14.4035 & \cellcolor{red!20} 14.9035 & \cellcolor{red!20} 14.4035
  \\ \cline{2-6}
     & LSCII & \cellcolor{green!22} 3.00415 & \cellcolor{red!20} 14.2541 & \cellcolor{red!20} 14.2541 & \cellcolor{red!20} 14.2541
  \\ \hline\hline
  \multirow{2}{*}{3} & TT-TFD & \cellcolor{yellow!40} 9.5034 & \cellcolor{red!20} 13.7534 & \cellcolor{red!20} 13.5034 & \cellcolor{red!20} 14.0034
  \\ \cline{2-6}
     & LSCII & \cellcolor{green!22} 7.25415 & \cellcolor{yellow!40} 9.25415 & \cellcolor{red!20} 12.0041 & \cellcolor{yellow!40} 11.5041
  \\ \hline\hline
  \multirow{2}{*}{4} & TT-TFD & \cellcolor{red!20} 14.6535 & \cellcolor{green!22} 5.65348 & \cellcolor{red!20} 14.9035 & \cellcolor{yellow!40} 11.9035
  \\ \cline{2-6}
     & LSCII & \cellcolor{green!22} 8.50415 & \cellcolor{green!22} 6.00415 & \cellcolor{red!20} 13.5041 & \cellcolor{yellow!40} 11.7541
  \\ \hline\hline
  \multirow{2}{*}{6} & TT-TFD & \cellcolor{yellow!40} 9.65348 & \cellcolor{yellow!40} 10.4035 & \cellcolor{red!20} 13.9035 & \cellcolor{red!20} 13.6535
  \\ \cline{2-6}
     & LSCII & \cellcolor{red!20} 12.7541 & \cellcolor{red!20} 14.7541 & \cellcolor{red!20} 14.2541 & \cellcolor{red!20} 13.5041
  \\ \hline
\end{tabular} 
}
\label{tab:memtime}
\end{table}

The computational cost of the GQME approaches with respect to dimensionality therefore depends on several factors with different and at times opposing scaling trends. Thus, the computational cost benefits of reduced-dimensionality GQMEs depends on the balance between these trends and further work is needed to determine whether using a reduced-dimensionality GQME provides a way to significantly reduce computational cost.


\section{Concluding Remarks}
\label{sec:conc}

We have implemented the TT-TFD method to obtain quantum-mechanically exact memory kernels and inhomogeneous terms for different types of GQMEs describing the dynamics of electronic DOF for the spin-boson model. We have analyzed a GQME for the 4-element full electronic reduced density matrix, a populations-only GQME for the two diagonal elements, and single-population scalar GQMEs for single diagonal elements. We have also demonstrated that all four GQMEs are exact equations of motion and thus reproduce the same exact population dynamics when parametrized by a quantum-mechanically exact input method such as TT-TFD, although the four GQMEs are different in form and dimensionality. 

Advancing the capability to calculate quantum-mechanically exact memory kernels and inhomogeneous terms for different types of GQMEs is highly desirable for multiple reasons:
\begin{itemize}
\item First, it should be noted that the memory kernels and inhomogeneous terms in the case of quantum open systems serve a similar role to that of the Hamiltonian in the case of closed quantum systems. More specifically, similar to how analyzing the properties of the Hamiltonian is often used to shed light on the closed quantum system dynamics it gives rise to, one expects that knowing what the quantum-mechanically exact memory kernel and inhomogeneous term look like and how they depend on various parameters and different choices of projections could shed light on the open quantum system dynamics they give rise to. 
\item Second, quantum-mechanically exact memory kernels and inhomogeneous terms are particularly valuable to evaluate the capabilities of PFIs obtained with approximate input methods as shown in our comparisons of memory kernels and inhomogeneous terms obtained with exact TT-TFD and approximate LSCII simulation methods. 
\item Third, quantum-mechanically exact memory kernels can be used as benchmarks to assess the accuracy of various types of perturbative 
quantum master equations (QMEs).\cite{redfield57,pollard94,pollard96,meier99,egorova01,baiz11,laird91,zhang98,novoderezhkin04,trushechkin19,jang20,lai21}  More specifically, while the GQMEs correspond to the exact equations of motion of the subset of dynamical quantities of interest, the derivation of perturbative QMEs relies on approximate expressions for the memory kernels that are based on treating various terms in the Hamiltonian, such as the system-bath coupling or electronic coupling, as small perturbations. 
Thus, comparisons of the perturbative memory kernels to the exact kernels can provide a better understanding of the accuracy of perturbative methods and their range of validity. 
\item Fourth, in certain situations, simulating the quantum dynamics via a GQME may be more cost-effective than the direct use of the numerically-exact quantum dynamics method. More specifically, restricting the use of a quantum-mechanically exact method to calculating the 
PFIs can provide a more efficient route to obtain the dynamics of the quantity of interest compared to extracting it from the overall system dynamics. The computational cost analysis of the TT-TFD-based GQME approach provided in this paper constitutes the first step towards understanding when and how simulating the quantum dynamics via a GQME approach is advantageous compared to the direct use of the numerically-exact quantum dynamics method. 
\end{itemize}

Various extensions of this study would be highly desirable, including combining the GQME approach with other quantum-mechanically exact and approximate input methods, calculating memory kernels and inhomogeneous terms for other types of dynamical quantities of interest and exploring the capabilities of the GQMEs on other benchmark models. 
Work on such extensions is currently underway and will be reported in future publications.

\section*{Acknowledgments}

EG and VSB acknowledge support from the NSF grant 2124511 [CCI Phase I: NSF Center for Quantum Dynamics on Modular Quantum Devices (CQD-MQD)]. MBS acknowledges support from the Yale Quantum Institute Postdoctoral Fellowship.  Also acknowledged are computational resources and services provided by the Advanced Research Computing at the University of Michigan, Ann Arbor. We thank Paul Bergold for stimulating discussions.


\section*{Data Availability}

The code for the TT-TFD + GQME simulation of Model 1 is available at \href{https://github.com/NingyiLyu/TT-TFD-GQME}{https://github.com/NingyiLyu/TT-TFD-GQME}. The data that supports the findings of this study are available within the article and SI.


\appendix


\bibliography{bib}

\includepdf[pages=-]{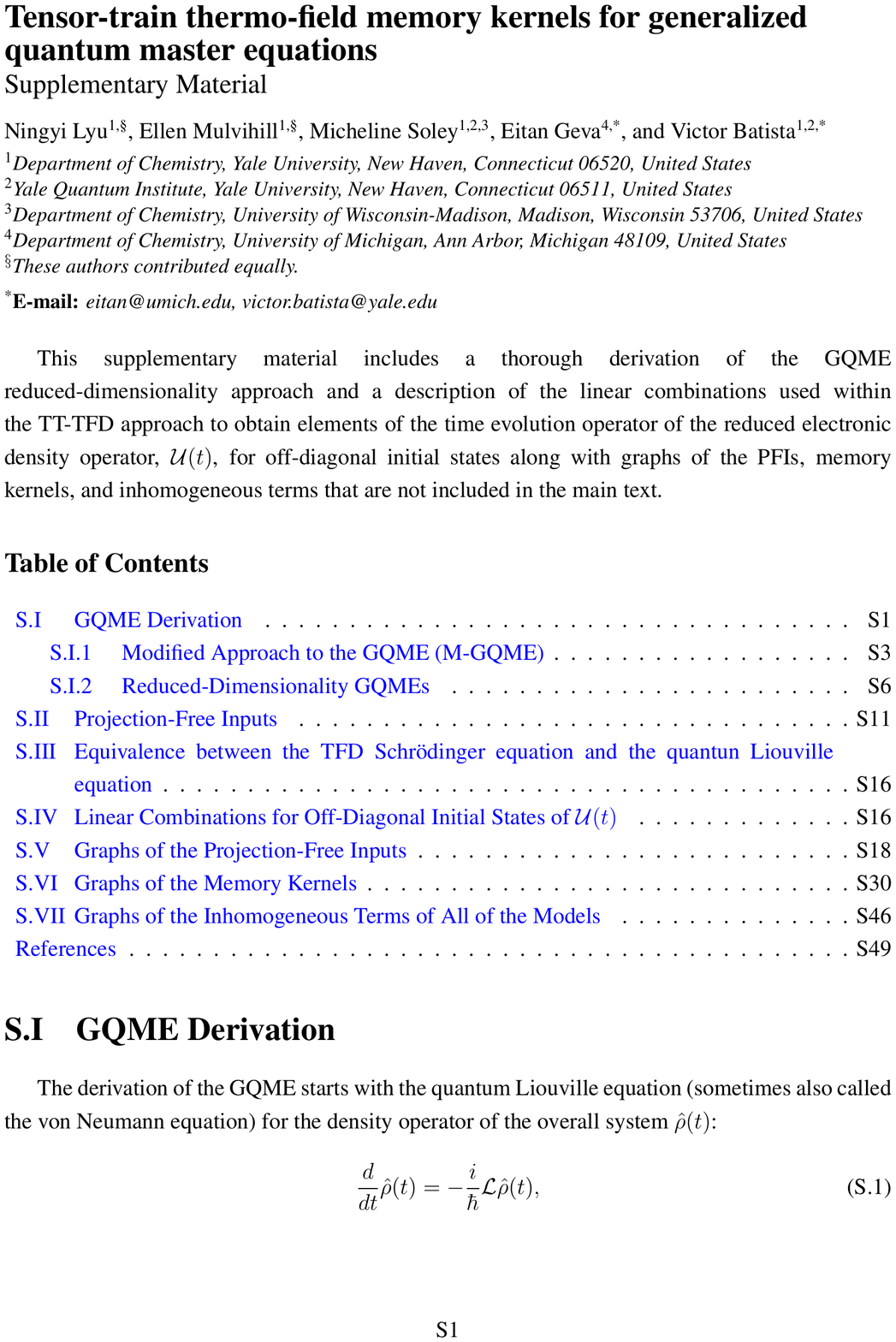}
\end{document}